\documentclass[12pt]{iopart}
\usepackage{graphicx}
\expandafter\let\csname equation*\endcsname\relax
\expandafter\let\csname endequation*\endcsname\relax
\usepackage{amsmath}

\begin{document}

\title{Skyrmions Near Defects}

\author{Amel Derras-Chouk }

\author{Eugene M. Chudnovsky}
\address{Physics Department, Herbert H. Lehman College and Graduate School,
The City University of New York \\
 250 Bedford Park Boulevard West, Bronx, New York 10468-1589, USA}

\date{\today}

\begin{abstract}
We study the impact of an exchange-reducing defect on a skyrmion in a thin film of finite thickness. Attraction of the skyrmion to a defect is demonstrated in a lattice model by computing the micromagnetic energy accounting for the exchange, Dzyaloshinskii-Moriya interaction, magnetic anisotropy, and dipole-dipole coupling. The spiraling dynamics of the skyrmion towards the defect is illustrated by solving numerically the full Landau-Lifshitz-Gilbert equations on a lattice and, independently, by solving the Thiele equation, with the two methods in agreement with each other. Depinning of the skyrmion by the current is investigated. We find that the skyrmion deforms when it is close to the defect. Deformation is small in the parameter space far from the phase boundary that determines stability of skyrmions. It increases dramatically near the phase boundary, leading to the transformation of the skyrmion by the defect into a snake-like magnetic domain. 
\end{abstract}

\maketitle

\section{Introduction}

The interplay of interactions in thin magnetic films can lead to the formation of skyrmions, vortex-shaped spin textures that are topologically stable in the continuous spin-field limit. They may be manipulated by electric currents that are much lower than the currents needed to move domain walls \cite{Fert2017,Kang2016b,Leonov2015}. Skyrmions can be written on a magnetic film and moved along a track, making them good candidates for energy efficient memory storage \cite{Fert2017, Kang2016b, Leonov2015, Romming2013}.

Many materials that host skyrmions possess Dzyaloshinskii-Moriya interaction (DMI). The interaction emerges from the spin-orbit coupling in lattices that lack inversion symmetry and can be induced at the interface of magnetic layers and heavy metals. Skyrmions lattices were originally found in bulk helical magnets by Yu et  al \cite{Yu2010}. They were later observed in thin films made up of PdFe bilayers on Ir(111), a setup leading to significant DMI contribution \cite{Heinze2011}. These skyrmions live at the interface of the materials, effectively constrained to two dimensions. However, skyrmions can also exist in materials that do not contain DMI, as stabilizing effects can emerge from other interactions \cite{Fert2017, Kang2016b}. In either case, the exchange interaction is the dominant term in the energy contributing to skyrmion stability. 

In practice, a skyrmion can often come into close proximity of a crystal defect, making the effect of defects an important issue for applications of skyrmions. The behavior of a skyrmion near a defect that modifies the exchange interaction has been studied by a number of authors \cite{Liu2013, Lin2013, Navau2018, Stosic2017, LimaFernandes2018, Hanneken2016}. Some examined in-layer atomic defects in thin films \cite{Hanneken2016, Lin2013, Liu2013}, and others considered in-plane linear defects \cite{Navau2018, Stosic2017}. While some of these studies showed consistent attraction to a point defect \cite{Liu2013,Lin2013}, others demonstrate that the final equilibrium position of a skyrmion near an exchange modifying defect is a nontrivial function of material parameters \cite{Hanneken2016, Stosic2017,LimaFernandes2018}. The exact form of the point defect also influences the results \cite{Navau2018}. 

Previous work has additionally explored the impact of a defect on the DMI and anisotropy energies \cite{Hanneken2016, Stosic2017, Navau2018}. Stosic et al. showed that the equilibrium position of a skyrmion along a linear defect varies based on the type of the defect, revealing potential ways to control skyrmion direction in a racetrack memory \cite{Stosic2017}. Navau et al. described a more complicated picture where attraction or repulsion from in-plane linear DMI or anisotropy defects depends on the distance to the skyrmion among other variables \cite{Navau2018}.

Experimental work by Hanneken et al. \cite{Hanneken2016} in PdFe/Ir(111) used scanning tunneling microscopy to observe that skyrmions become pinned near in-layer atomic defects, suggesting that the skyrmions are more stable near defects introduced in the Pd layer. This has been in line with the studies of other authors \cite{Pulecio2016, Zeissler2017,Arjana2020}.  An extended quenched disorder due to the distribution of defects has been shown to have a significant effect on skyrmion phases \cite{Garanin2020}. Additionally the movement of skyrmions affected by disorder that originates from grain boundaries has been investigated  \cite{Salimath2019, Litzius2020}. 

Lin et al. have examined temperature-induced skyrmion motion described by the Thiele equation that allows a skyrmion, treated as a particle, to escape from the potential well created by the modification of the exchange interaction \cite{Lin2013}. The depinning of skyrmions from defects driven by a spin-polarized current has been studied \cite{Lin2013,Liu2013, Muller2015}. Recently, a quantum depinning mechanism has been suggested \cite{Psaroudaki2020}. 

Much of the previous research shows that defects cause skyrmions to deform \cite{Kim2017, Hanneken2016, LimaFernandes2018}. While most analytical theories of skyrmions assume cylindrical symmetry \cite{Kang2016b, Buttner2018}, skyrmion deformation near a defect illustrates that analytical models should be adjusted to reflect skyrmion deformation away from cylindrical symmetry. It is especially important when the external field is low and the system is close to the phase boundary that separates skyrmion states from labyrinth domains. This has been noted in the work exploring the impact that defects have on skyrmion lattices \cite{Pierobon2018}. It was shown that the presence of even a single defect can turn an otherwise stable skyrmion lattice into a domain state.

It is this final note regarding deformation of skyrmions by defects that our paper explores more directly in systems containing a single skyrmion rather than a skyrmion lattice. This choice is motivated by the fact that applications of skyrmions  require manipulation of individual skyrmions. We use both micromagnetic computations on a lattice and functional minimization in the continuum model to examine the impact that an exchange-reducing defect has on a skyrmion. Numerically finding that the skyrmion deforms as it approaches the defect prompted us to develop a semi-analytical method that allows such deformations. This includes a correction to the standard cylindrically-symmetric spin texture considered in most works and allows for distortions of the skyrmion to find new energy minima in the phase space. 

The paper is structured as follows. Section \ref{skyrmion-defect-interaction} gives an overview of the type of defect we consider and the numerical methods used in the micromagnetic computation. It also describes the method used to find the local minimum in the phase space that defines the skyrmion configuration in the continuous limit. Details of both numerical methods are in the Appendix. Section \ref{skyrmion-dynamics} presents results of skyrmion dynamics near a defect and comparison with the dynamics described by the Thiele equation. Section \ref{skyrmion-depinning} presents results of current-driven skyrmion motion near a defect. Section \ref{skyrmion-deformation} illustrates that an otherwise stable skyrmion may become unstable against transformation into a snake-like magnetic domain near a defect. Our results are discussed in Section \ref{conclusion}.

\section{Skyrmion-Defect Interaction}\label{skyrmion-defect-interaction}

\subsection{Micromagnetic Approach}

We consider a skyrmion stabilized by the exchange, Zeeman, Dzyaloshinskii-Moriya, anisotropy, and dipole-dipole interactions. The total energy of the thin film is given by

\begin{align}\label{energy-lattice}
\mathcal{H}  =  &-\frac{1}{2}\sum_{i,j} J_{ij} \,\, \mathbf{s}_i \cdot \mathbf{s}_j - H \sum_i s_{z,i} \nonumber \\ 
&+ A \sum_i \bigg(  (\mathbf{s}_i  \times \mathbf{s}_{i+\hat{x}})\cdot \hat{x} + (\mathbf{s}_i \times \mathbf{s}_{i+\hat{y}})\cdot \hat{y} \bigg)  \nonumber \\
 & - \frac{D_z}{2} \sum_{i} s_{z,i}^2 - \frac{E_D}{2}\sum_{i,j} \Phi_{ij,\mu\nu} s_{i\mu} s_{j\nu}.
\end{align}

\noindent Here $J_{ij}$ is the constant of the exchange interaction between spins $i$ and $j$, $H=g\mu_{B}SB$ is the external field in energy units, where $g$ is the gyromagnetic ratio and $\mu_{B}$ is the Bohr magneton. Constant $A$ describes the strength of the DMI. Spin $\mathbf{s}_i$ is a three component vector of magnitude $S$ at the lattice point $i$. The exchange sum is taken over the nearest-neighbor spins. The last two terms refer to the perpendicular magnetic anisotropy (PMA) and dipole-dipole interaction (DDI), with $D_z$ being the strength of the PMA, $E_D= \mu_0 M_0^2 a^3 /(4\pi)$ being the DDI constant, and $M_0=g\mu_B S/a^3$ being the magnetization. Here $a$ is the lattice spacing. The DDI term in Eq. (\ref{energy-lattice}) contains tensor

\begin{align}\label{ddi-eqn}
\Phi_{ij,\mu\nu} = \frac{a^3}{r_{ij}^{5}}(3r_{ij,\mu}r_{ij,\nu} - \delta_{\mu\nu}r^2_{ij}),
\end{align}

\noindent where $r_{ij}=r_i - r_j$ is the distance between the lattice sites $i$ and $j$, and $\mu, \nu$ refer to the Cartesian components $x, y,z$, and $\delta_{\mu,\nu}$ is the Kronecker unit tensor. 

Since the DDI is long-range, its inclusion in the energy lengthens the numerical computation considerably. We make large-size micromagnetic computations tractable by considering a thin film with spin directions nearly constant along the thickness of the film. (Details of the numerical method can be found in Ref. \cite{Capic2019}.) We use the term thin-film in reference to the thickness of a magnetic nanolayer deposited on a heavy-metal substrate. This makes the problem effectively two dimensional by allowing one to consider the interaction between columns of spins instead of individual spins.

Computations presented in this paper study skyrmions in thin films of thickness $N_z = 10$. In considering materials containing both DDI and PMA, it is useful to define the ratio between the two interaction strengths as $\beta = D_z/(4\pi E_D)$. We choose a practically reasonable ratio of $\beta = 1$. All computations used a positive DMI constant and a negative external field in the direction of ferromagnetically aligned spins far from the center of the skyrmion. These choices, along with the form of the DMI energy used in Eq.\ (\ref{energy-lattice}), lead to the formation of a Bloch-type skyrmion with a central spin pointing opposite to the direction of the external field. Studies presented in this paper can be repeated with Neel-type skyrmions by replacing the DMI term with the appropriate Neel analog.

With the energy of the system defined, we introduce a defect into the lattice. The type of defect we consider modifies the exchange energy according to

\begin{align}\label{modified-exchange}
J_{ij} = J_{0}\bigg( 1 + \alpha \exp\bigg( -\frac{|(\mathbf{r}_{i} + \mathbf{r}_j)/2 - \mathbf{r}_d|^2}{\xi^2} \bigg) \bigg).
\end{align} 

\begin{figure}
\centering
\includegraphics[width=0.6\linewidth]{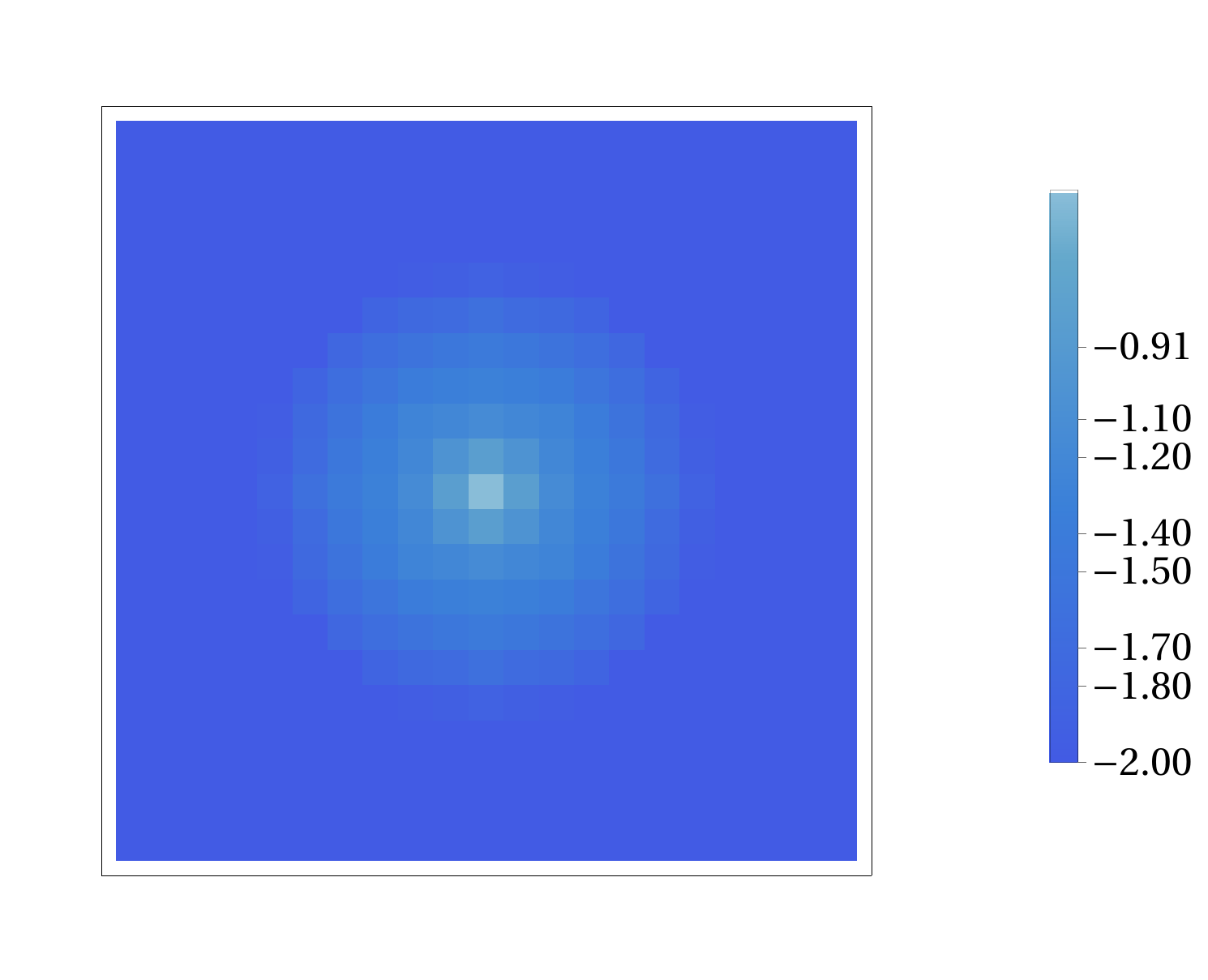}
\caption{The exchange energy density of a spin lattice with a defect placed at the center is reduced near the defect, as shown in the figure. The defect parameters are $\alpha=-0.7$ and $\xi=1.0$.}
\label{fig:exchange-energy-density}
\end{figure}

\noindent Here, $J_0$ is the exchange constant at infinity, which is set to 1, $\alpha$ defines the strength of the exchange modification, $r_d = (x_d, y_d)$ denotes position of the defect, and $\xi$ defines the width of exchange modification. The other position vectors, $\mathbf{r}_i$ and $\mathbf{r}_j$, specify positions of lattice sites. The distance of the midpoint of lattice sites from the center of the defect is used to quantify the reduction in exchange energy between lattice sites $i$ and $j$. Fig. \ref{fig:exchange-energy-density} shows the exchange energy density for a $256 \times 256$ lattice hosting a skyrmion. Similar Gaussian-type defects have been considered in works that examine skyrmion-defect interaction \cite{Hanneken2016,Liu2013}. In a $2D$ system Eq. (\ref{modified-exchange}) describes the effect of a point atomic defect, like a vacancy or an interstitial atom. In a film of a finite thickness it models the modification of the exchange interaction by a linear dislocation going in the $z$ direction. Such dislocations are common in thin films and their effect on skyrmions must be much stronger than the effect of point atomic defects. 

In order to determine the energy of the interaction between the skyrmion and the defect, we place the defect at the center of the lattice, setting $x_d=0$ and $y_d=0$, and pin the skyrmion at a distance $\Delta/a$ away from the center. Then we compute the minimum energy configuration of the system for various pinning distances, ultimately computing the energy as a function of the distance between the skyrmion and the defect. There are several ways to numerically determine the minimum energy configuration of a set of spins in a lattice, many of which involve finding a local minimum in the energy phase space. We follow this formula and determine minimum energy configurations by evaluating the Landau-Lifshitz-Gilbert equation in time. Details of the numerical implementation are given in \ref{appendix-micromagnetic}.

\begin{figure}
\centering
\includegraphics[width=0.75\linewidth]{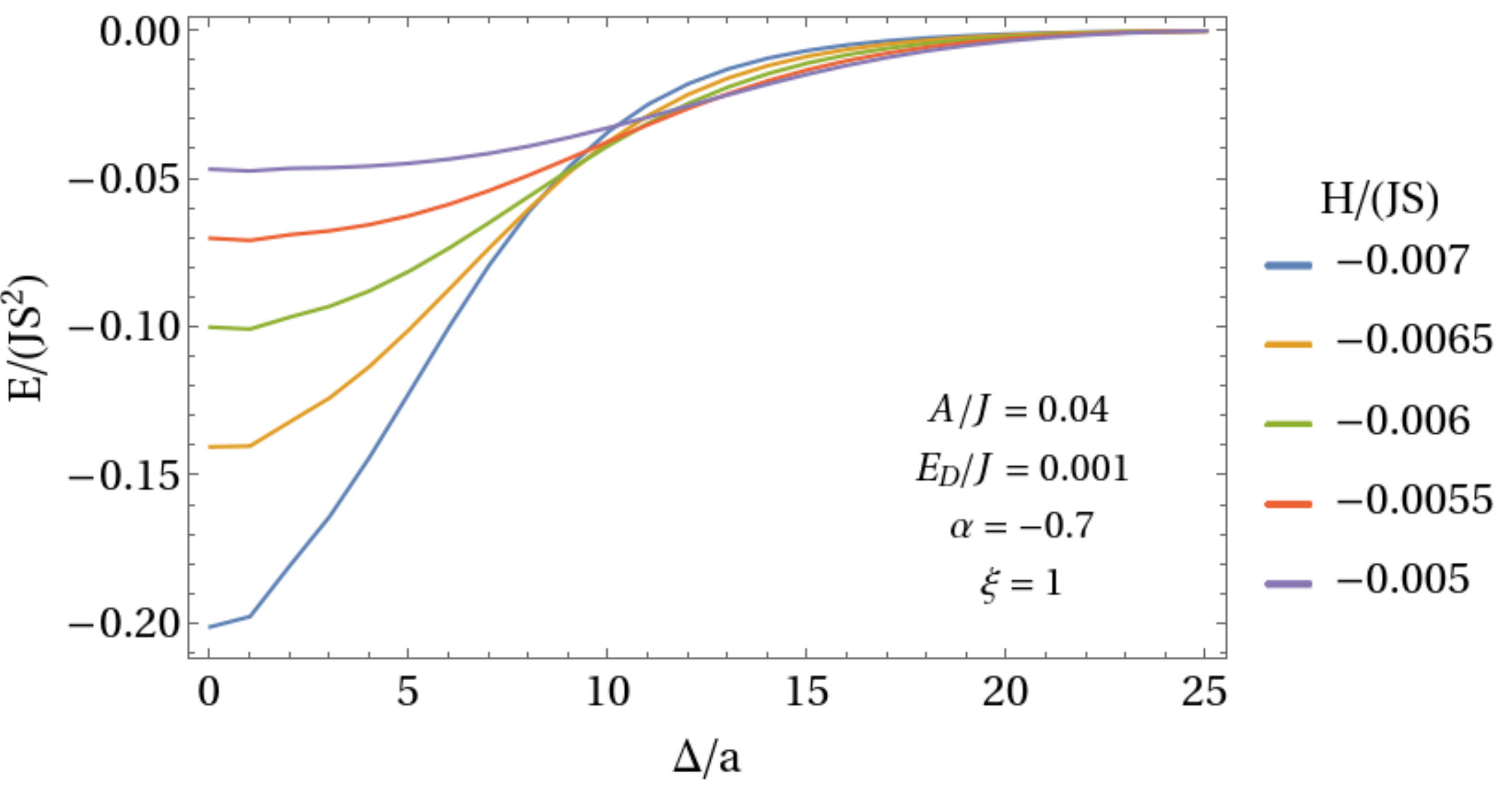}
\caption{Energy of the skyrmion as a function of the distance between the defect and the skyrmion. The defect is located at the center of the lattice and the skyrmion is pinned at various distances, $\Delta/a$, along the $y$ axis.}
\label{fig:energy-vs-distance}
\end{figure}

Results show that the energy is lower when the center of the skyrmion coincides with the exchange-reducing defect. This is expected, as the exchange energy of a skyrmion, which is the dominant part of its energy, is maximal at the center. Fig. \ref{fig:energy-vs-distance} illustrates the decrease in the energy as the skyrmion nears the defect for material parameters $A/J = 0.04$, $E_{D}/J=0.001$, and various magnetic fields.

\subsection{Functional Approach}

The behavior of a skyrmion near a defect can also be explored in the continuous energy model. To do so we define the total energy as 

\begin{align}\label{energy-integral-xy}
E =\int_{-\infty}^{\infty} \int_{-\infty}^{\infty} & dx dy \,  \bigg(  a^4 \frac{J_{ex}}{2} \bigg[ \big(\partial_x \textbf{n}\big)^2+\big(\partial_y \textbf{n} \big)^2 \bigg] \nonumber \\
& + a^3 A \bigg[ (\textbf{n}\times \partial_x \textbf{n}) \cdot e_x + (\textbf{n}\times \partial_y \textbf{n}) \cdot e_y \bigg] \nonumber \\
& - H (n_z-1) - D_z (n_z^2-1) \bigg).
\end{align}

\noindent Here, $a$ is the lattice constant and $\textbf{n} = \textbf{S}/S$ where $S$ is the spin field density, $\tilde{S}/a^2$. The other parameters, $J_{ex}$, $A$, $H$, and $D_z$ refer to the exchange, DMI, Zeeman, and PMA constants. In the final two terms we subtract $1$ from $n_z$ so that expression can be integrated over all space. The resulting expression is the total energy relative to a ferromagnetic state. We include DDI as an effective anisotropy, which is permitted when the skyrmion is smaller than the film thickness \cite{Bogdanov1994}. For skyrmions much larger than the film thickness, the DDI decays exponentially and can be omitted entirely \cite{Leonov2015}. This model is used to examine skyrmion behavior near a defect as was done in the micromagnetic model, though exact quantitative agreement between the models is not expected because we do not account for lattice effects in the continuous model.

We seek a spin configuration $\textbf{n}(x,y)$ which will minimize the integral of Eq. (\ref{energy-integral-xy}). The problem is simplified by assuming a cylindrically symmetric spin texture given by  

\begin{align}\label{spin-general}
\textbf{n} = \{\sin[f(r)]\cos[g(\phi)],  \sin[f(r)]\sin[g(\phi)], \cos[f(r)]\}.
\end{align}

\noindent Here $f(r)$ and $g(\phi)$ are functions that define the shape of the spin texture. One can show that for a skyrmion of topological charge 1, $g(\phi) = \phi + \gamma$, where $\gamma$ is the skyrmion chirality. We are interested in a Bloch skyrmion with $\gamma=\pi/2$. Substituting this into the integral of Eq. (\ref{energy-integral-xy}) and converting from Cartesian to cylindrical coordinates gives

\begin{align}\label{energy-cylindrical}
E = & \int_0^{\infty} \int_0^{2\pi}  r \, dr \, d\phi  \bigg\{ \, a^4 J_{ex} \bigg[ \frac{1}{2}f_{r}(r)^2 + \frac{\sin^2(f(r))}{2r^2}\bigg] \nonumber \\
& + a^3 A\bigg[ f_{r}(r) + \frac{\sin(2f(r))}{2r} \bigg] - H [\cos(f(r))-1] \nonumber \\
&- D_z [\cos(f(r))-1]^2 \bigg\}.
\end{align}

The function $f(r)$ that minimizes this integral can be found by either solving the respective Euler-Lagrange equation or optimizing the integral directly. We use the latter method to find minimum energy solutions by inserting an ansatz $f = f_0(r) = \pi \exp(-a r) \text{sech}(b r)$ and using local optimization tools to determine the parameters in the ansatz that will minimize the energy. 

\begin{figure}
\centering
\includegraphics[width=0.65\linewidth]{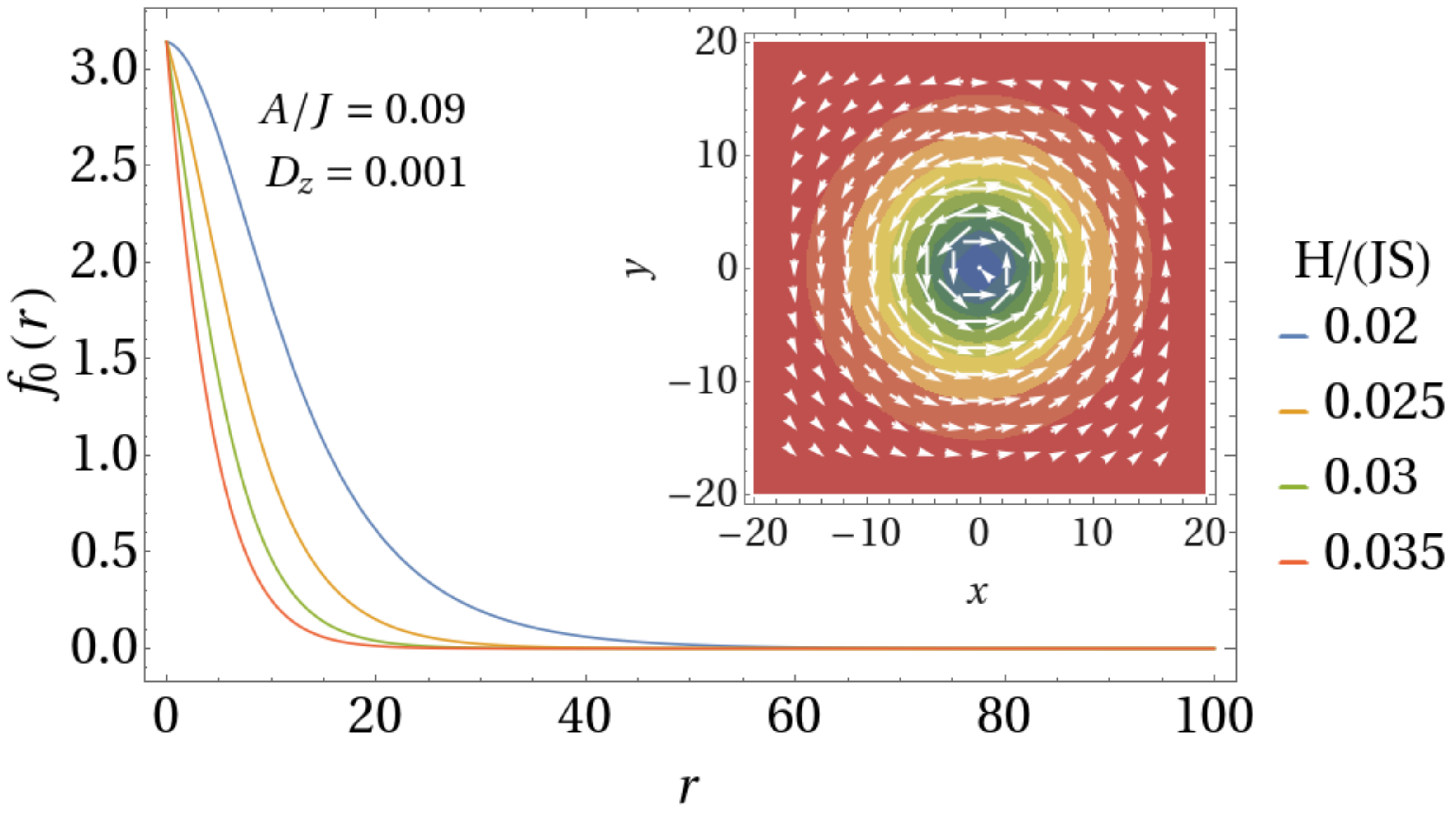}
\caption{Various $f_0(r)$ curves show the minimum energy solutions determined from minimization of the energy functional for different external fields. Inset shows the skyrmion in the Cartesian plane where the arrows indicate $x$ and $y$ components of $\mathbf{n}$ and the colors show magnitude of the z-component. Blue indicates $n_z$ into the page, while red indicates $n_z$ out of the page. }
\label{fig:f-of-r-no-defect}
\end{figure}

We consider small perturbations of the shape of a stable skyrmion that occur far from the phase boundary separating skyrmion states from domain states of the film. In this case, as we shall see below, deformations are small but finding them is instructive for the study of the interaction of the skyrmion with the defect. Deformations increase as one approaches the phase boundary, as seen in the next section. 

We begin by substituting the modified exchange interaction of Eq. (\ref{modified-exchange}) into the continuous integral and repeating the energy minimization. In the continuous model, unlike the micromagnetic model, we place the skyrmion at the center and move the defect. However, since we are interested in the impact of a defect on the nearby skyrmion, the problem no longer has cylindrical symmetry. Both $f$ and $g$ must be modified to account for this, which requires deriving Eq. (\ref{energy-cylindrical}) for non-centrosymmetric functions, and using a new ansatz in the energy minimization. Assuming that small changes in the energy landscape will cause the skyrmion to elongate, we perform the change of variables $\bar{r} = r + \epsilon \bar{f}(r) \bar{h}(\phi)$ and $\bar{\phi} = \phi + \epsilon \bar{g}(r,\phi)$, where $\epsilon$ is a small perturbation parameter. The trial functions are then expanded in the new coordinate system to the lowest order in $\epsilon$. We are interested in small elliptic distortions of $f_0(r)$, so we use $\bar{h}(\phi)=\cos(2\phi)$\cite{Bogdanov1994}. The functions $\bar{f}(r)$ and $\bar{g}(r,\phi)$ specify the radial strength of the distortion. Since we explore small perturbations, we assume both functions change slowly and are approximated with first-order Taylor-series expansions. 

\begin{figure}
\centering
\begin{minipage}{0.33\linewidth}
\includegraphics[width=\textwidth]{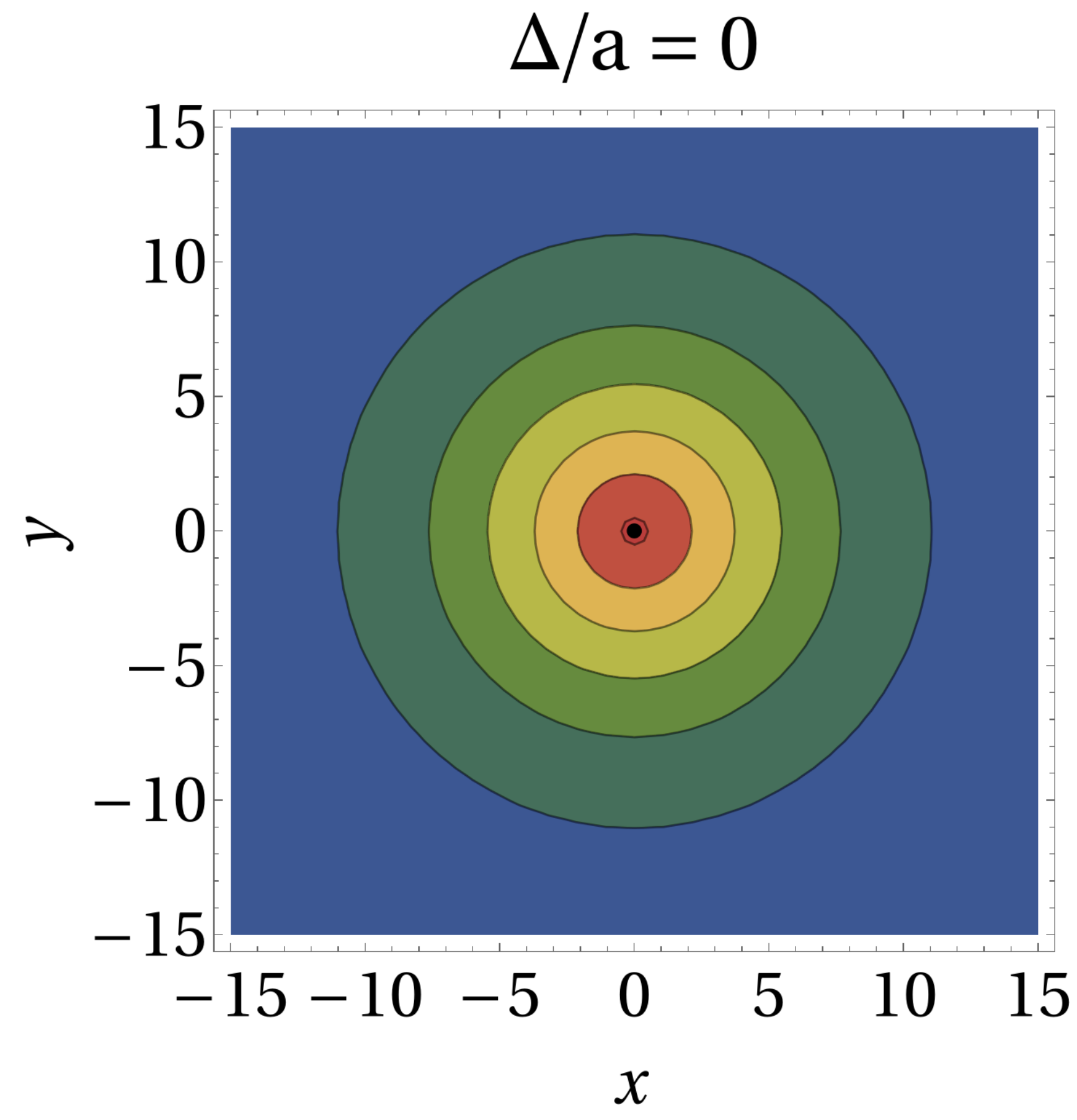}
\end{minipage}
\begin{minipage}{0.33\linewidth}
\includegraphics[width=\textwidth]{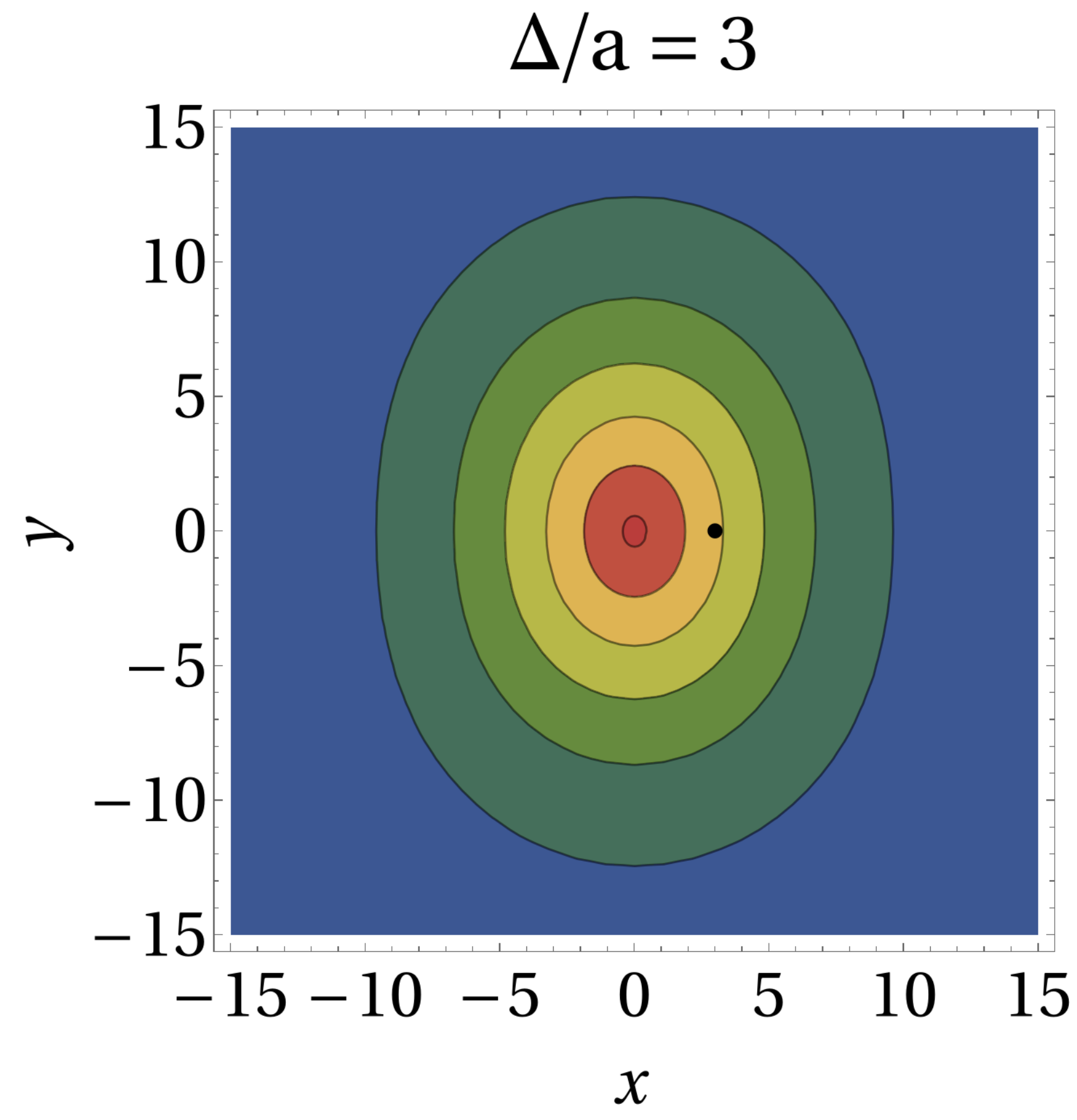}
\end{minipage}
\begin{minipage}{0.33\linewidth}
\includegraphics[width=\textwidth]{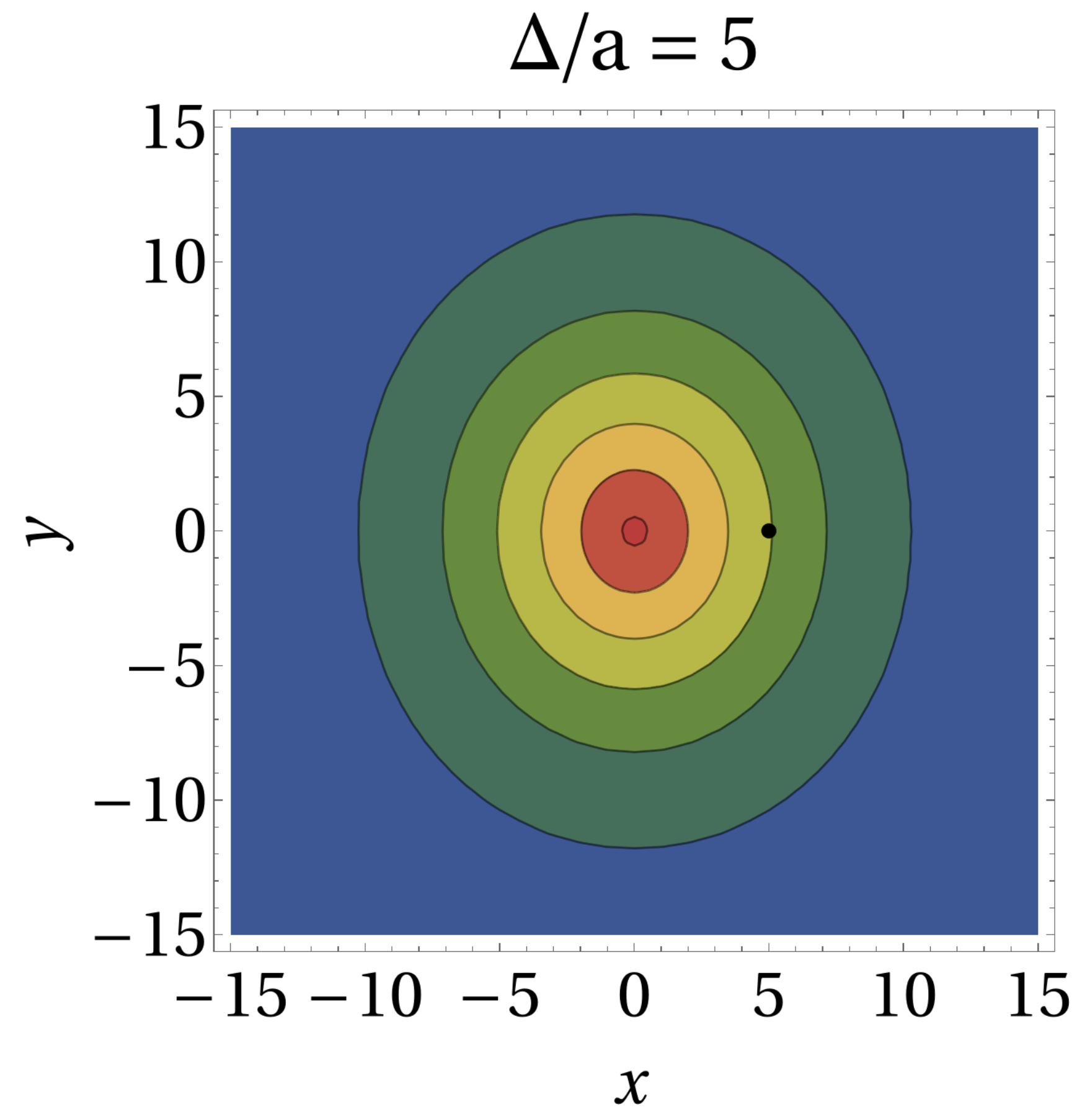}
\end{minipage}
\begin{minipage}{0.33\linewidth}
\includegraphics[width=\textwidth]{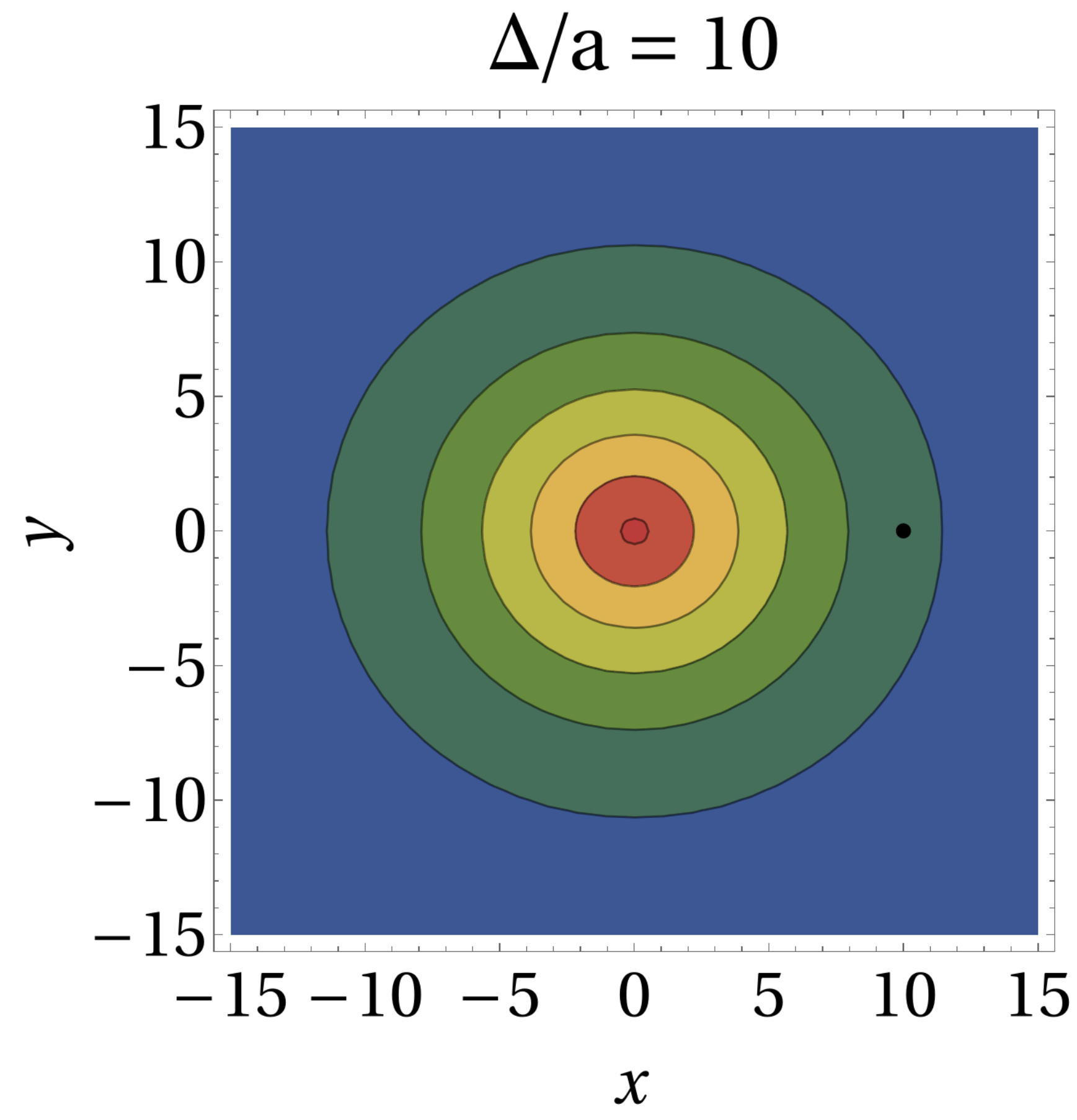}
\end{minipage}
\caption{Example of minimized $f(r)$ for a skyrmion in a lattice with a defect placed $\Delta/a$ away from the center of the skyrmion. Results are shown for $H/(JS)=0.03$, $A/J=0.1$, and $D_z/J=0.001$. Elongation has been magnified by a factor of 10 for better visibility. Functions $f(r)$ which minimize the total energy near a defect lead to skyrmion elongation when the defect is nearby. Contours show curves of constant $f(r)$, and colors show the function domain from zero (blue) to $\pi$ (red). Black point shows position of the defect.}
\label{fig:skyrmion-deform-energy-minimization}
\end{figure}

This method confirms our previous finding that the skyrmion energy is lower when the skyrmion is centered on a defect. It also shows that the skyrmion acquires elliptical shape as it approaches the defect, see Fig. \ref{fig:skyrmion-deform-energy-minimization}. Elongation of the skyrmion becomes apparent by plotting the minimum energy solutions for $f(r,\phi)$ in Cartesian space. The axis of the elongation depends on $\Delta/a$, with skyrmion approaching the defect first elongating vertically and then returning to the circular shape after centering on the defect. One can attempt to understand this intuitively by considering the orientation of spins in the skyrmion. It is advantageous for the skyrmion to reduce the exchange energy near its center since it is there that the misaligned spins lead to the increase of the exchange energy. When the skyrmion is close to the defect, it will attempt to increase the number of spins influenced by the exchange-reducing defect. The same is apparent when the skyrmion center is far from the defect and the skyrmion elongates to increase the contact between its edge and the defect.

\newpage

\section{Skyrmion Dynamics}\label{skyrmion-dynamics}

In this section we explore skyrmion dynamics near the exchange-reducing defect. The motion of a cylindrically symmetric skyrmion is governed by Thiele dynamics similar to that of a particle of charge $q$ moving in a magnetic field, with $q\mathbf{B} = G \hat{z}$ and $G = 4\pi \hbar Q/a^2$. Here $Q$ is the topological charge, which in our case is $1$. The equation of motion of the skyrmion is

\begin{align}\label{thiele-eqn}
\mathbf{G}\times \mathbf{\dot{\Delta}} + \mathbf{F} - K\mathbf{\dot{\Delta}} = 0.
\end{align} 

\noindent Here $K$ is the damping parameter, $\mathbf{F}$ is the force on the skyrmion, and $\Delta$ is the position of the skyrmion relative to the origin of the force. When there is no damping, i.e. $K=0$, the energy is conserved and the skyrmion moves in a circular path orbiting the source. 

The exchange-reducing defect modifies the energy of the skyrmion, $E(\Delta)$, and exerts a force, $F = -dE/d\Delta$, on the skyrmion. From the results shown in Fig. \ref{fig:energy-vs-distance} we can calculate this force directly and use Eq.\ (\ref{thiele-eqn}) to write for the skyrmion velocity, $v = \dot{\Delta}$,  at zero damping
\begin{align}
v = \frac{F}{G} = - \frac{1}{G}\frac{dE}{d\Delta}. 
\end{align}

\noindent A skyrmion that is $\Delta_0$ away from the defect has a frequency given by

\begin{align}\label{freq}
f = \frac{v}{2\pi \Delta_{0}} = \frac{F_{0}a^2}{ 8\pi^2 Q \hbar \Delta_0}.
\end{align}

\begin{figure}
\centering
\includegraphics[width=0.65\linewidth]{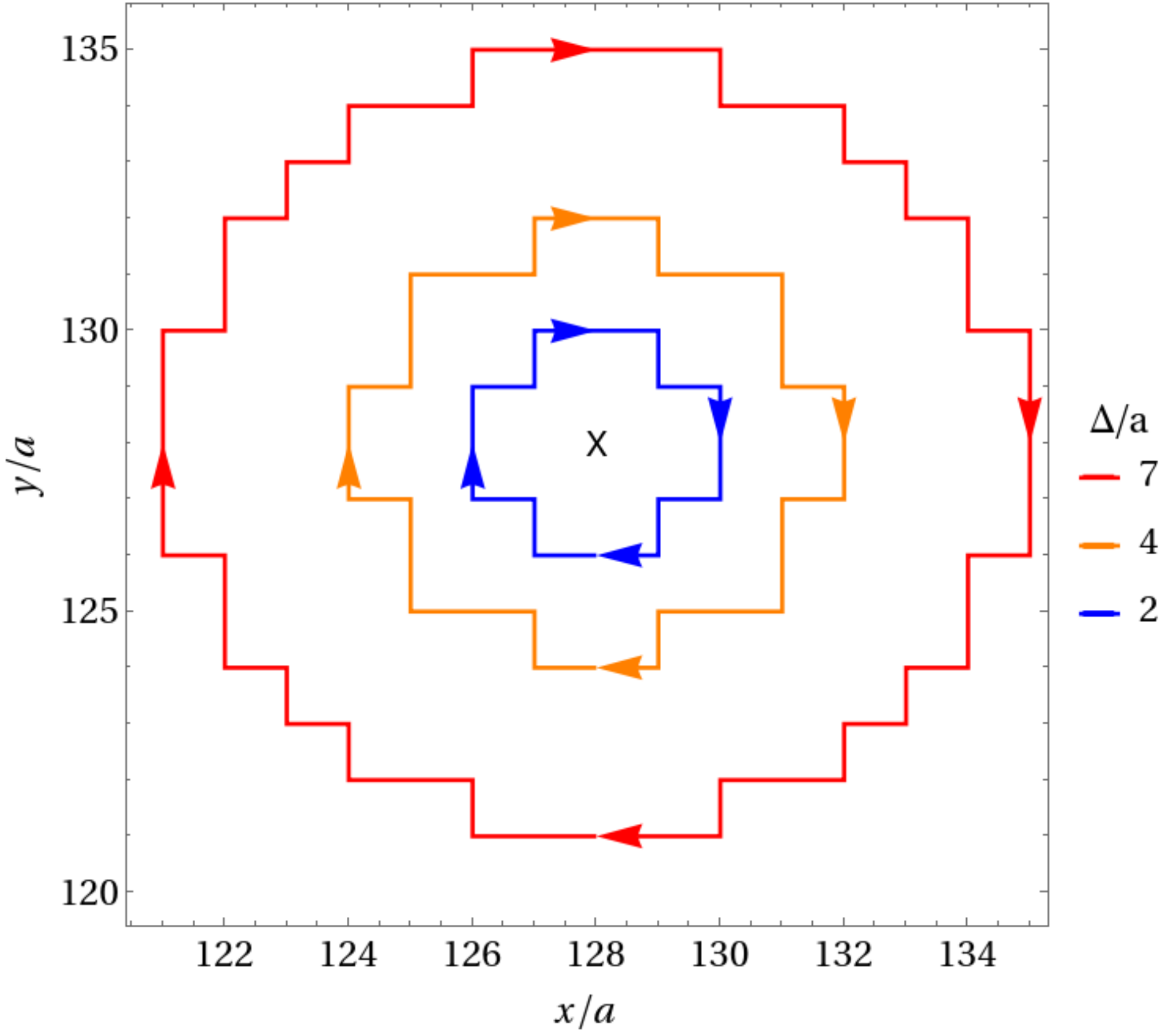}
\caption{Subsequent positions of the skyrmion for three distances from the defect in a 256 $\times$ 256 lattice, obtained by solving the LLG equations with zero damping. The lattice has been truncated to better observe the emerging trajectories. Results are shown for $H/(JS) = -0.0085$, $A/J=0.04$, and $E_D/J=0.001$.}
\label{fig:matplot-no-damping}
\end{figure}

\begin{figure}
\centering
\includegraphics[width=0.6\linewidth]{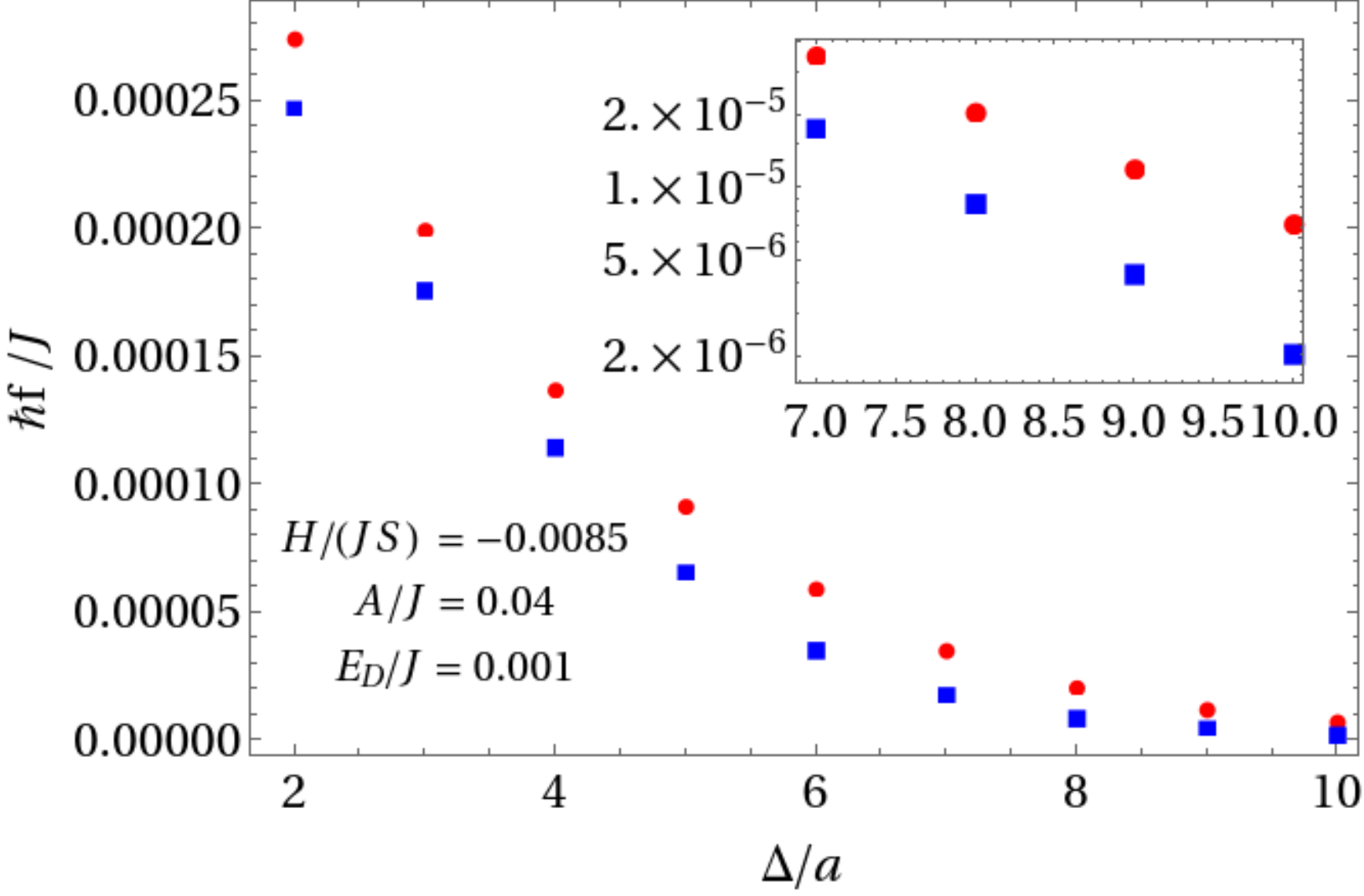}
\caption{Frequency of skyrmion rotation around the defect as a function of the radius. Blue points show the frequency determined from measuring the time it takes the skyrmion to rotate around the defect within the LLG dynamics with zero damping. Red points show the frequency determined by finding the force exerted on the skyrmion from the numerical results on the energy of its interaction with the defect and then using the Thiele equation to calculate frequency. Results are shown for a defect with $\alpha=-0.7$ and $\xi=1$.}
\label{fig:frequency-vs-radius}
\end{figure}

\noindent From Eq.\ (\ref{freq}) it is clear that the frequency of the rotation will decrease exponentially as the radius of skyrmion motion increases because the interaction energy between the skyrmion and the defect decreases exponentially. We use numerical results for $E(\Delta)$ to calculate the frequency of the rotation of the skyrmion in the Thiele approximation.

In addition we determine the frequency of the rotation from the full micromagnetic computation by computing evolution of the skyrmion spin texture according to the LLG equations of motion. The numerical procedure is described in \ref{appendix-micromagnetic}. In order to find the frequency of the rotation, we track the center of the skyrmion. This is done by tracing the position of the maximum value of $s_z$, since the center spin points in the direction opposite to that of the external field. Trajectories of the skyrmion at zero damping ($ \lambda = 0$) is shown in Fig. \ref{fig:matplot-no-damping} for three distances from the defect.

The frequency of the undamped rotation of the skyrmion around the defect, obtained from the Thiele dynamics and through micromagnetic computations, are shown in Fig. \ref{fig:frequency-vs-radius}. The two methods are in reasonable agreement with each other, though they increasingly differ as the skyrmion nears the defect. This is a reflection of skyrmion deformation as the skyrmion nears the defect, since the Thiele equation only applies to rigid, cylindrically symmetric skyrmions. Nevertheless, the reasonable agreement provides independent confirmation of our results for the energy of the skyrmion interaction with the defect.

\begin{figure}
\centering
\includegraphics[width=0.5\linewidth]{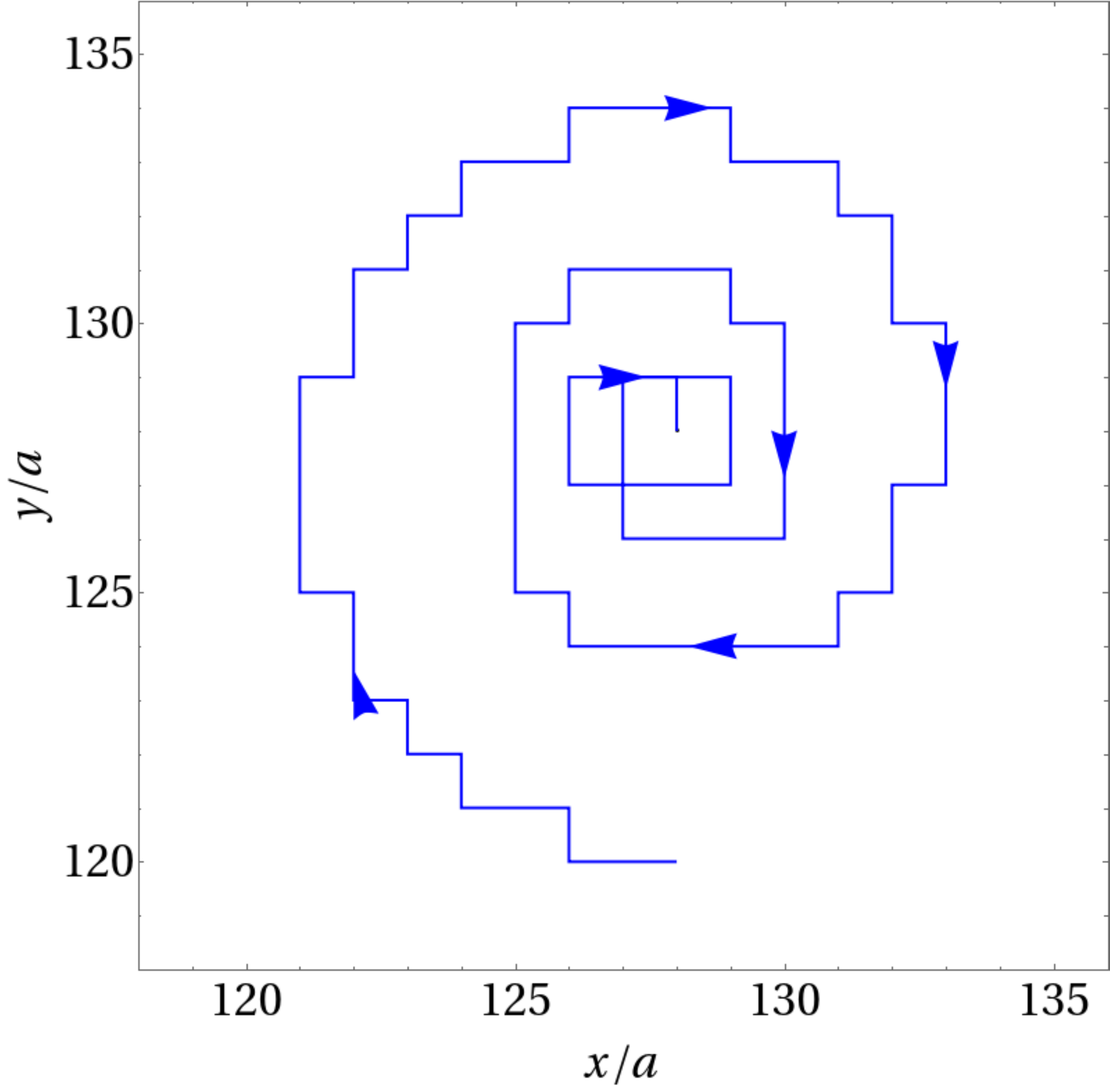}
\caption{Trajectory of the skyrmion computed in a 256 $\times$ 256 lattice obtained by solving the LLG equations with damping $\lambda = 0.1$ shows spiral motion toward the center of the defect. Results are shown for $H/(JS)=-0.0085$, $A/J=0.04$, and $E_D/J=0.001$.}
\label{fig:matplot-small-damping}
\end{figure}

\newpage

\section{Skyrmion Depinning}\label{skyrmion-depinning}

\begin{figure}[h!]
\centering
\begin{minipage}{0.7\linewidth}
\includegraphics[width=\textwidth]{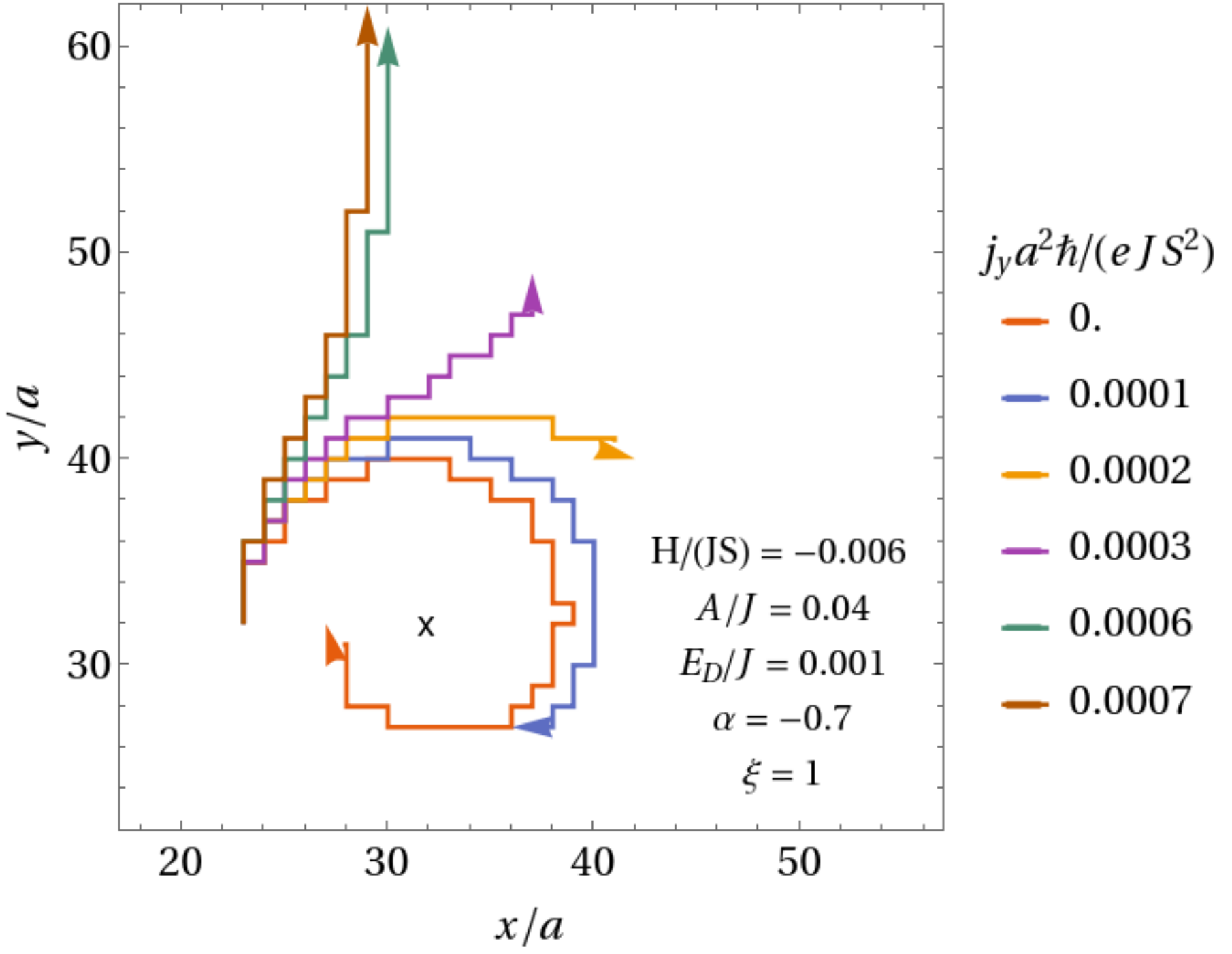}
\end{minipage}
\caption{Skyrmion position after $\hbar t/(JS^2)=10^{4}$ time steps on a $64\times64$ lattice. The skyrmion is initialized at a distance $10$ lattice sites from the defect whose location is marked with an `x'. A current is applied in the $y$ direction. If the current is below the critical value the skyrmion spirals towards the defect. Above the critical current the skyrmion moves away from the defect. }
\label{fig:add-to-paper}
\end{figure}

\begin{figure}
	\centering
	\includegraphics[width=0.55\textwidth]{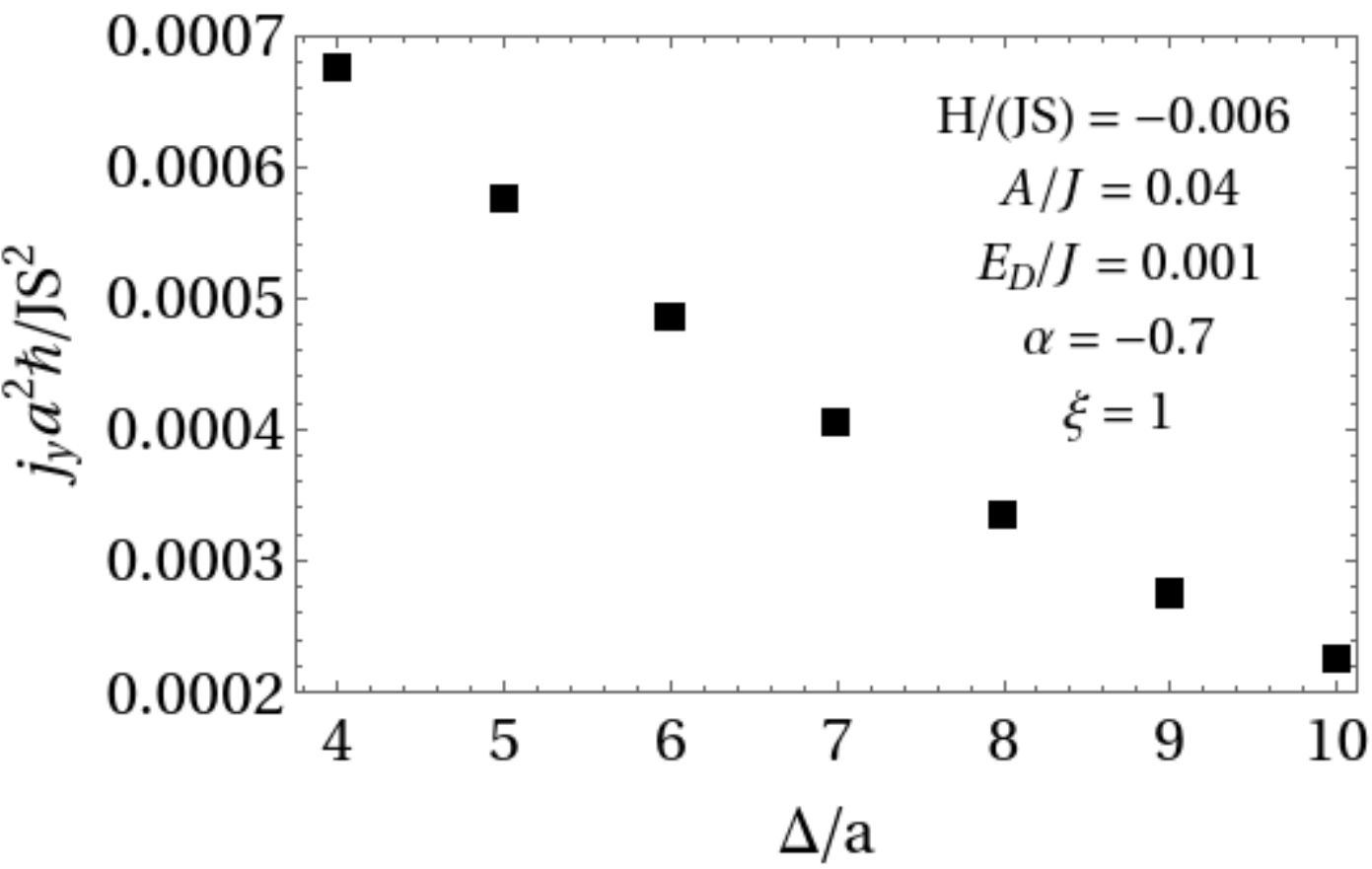}
	\caption{Critical current needed to drive the skyrmion away from the defect as function of skyrmion's initial distance from the defect.}
	\label{fig:energy-vs-time-current}
\end{figure}

Here we examine skyrmion depinning from an attracting defect by the spin-polarized current. Details of the numerical method are given in \ref{appendix-micromagnetic}. We measure the position of the skyrmion by tracking the location of the maximum $z$-component of the spin. Time dependence of the skyrmion location in the presence of a constant current for two different initial skyrmion positions are shown in Fig. \ref{fig:add-to-paper}. If the current is below the critical value the skyrmion spirals towards the defect. Above the critical current the skyrmion moves away from the defect. The current needed to keep driving the skyrmion away goes down as its distance to the defect increases. The dependence of the critical current on the distance of the skyrmion from the defect is shown in Fig. \ref{fig:energy-vs-time-current}. 

\newpage

\section{Skyrmion deformation near phase boundary}\label{skyrmion-deformation}

\begin{figure}
\centering
\begin{minipage}{0.45\linewidth}
\includegraphics[width=\textwidth]{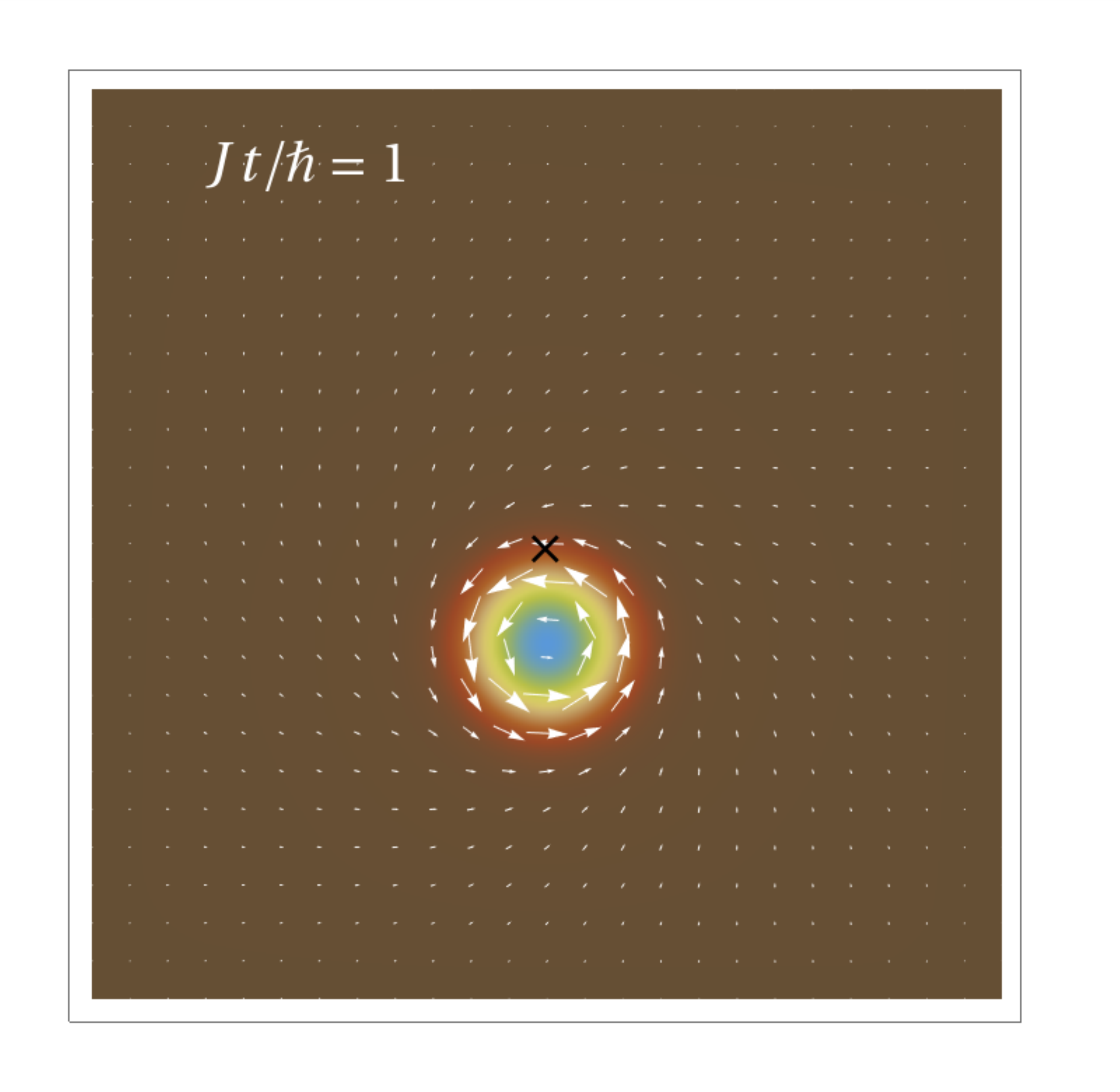}
\end{minipage}
\begin{minipage}{0.45\linewidth}
\includegraphics[width=\textwidth]{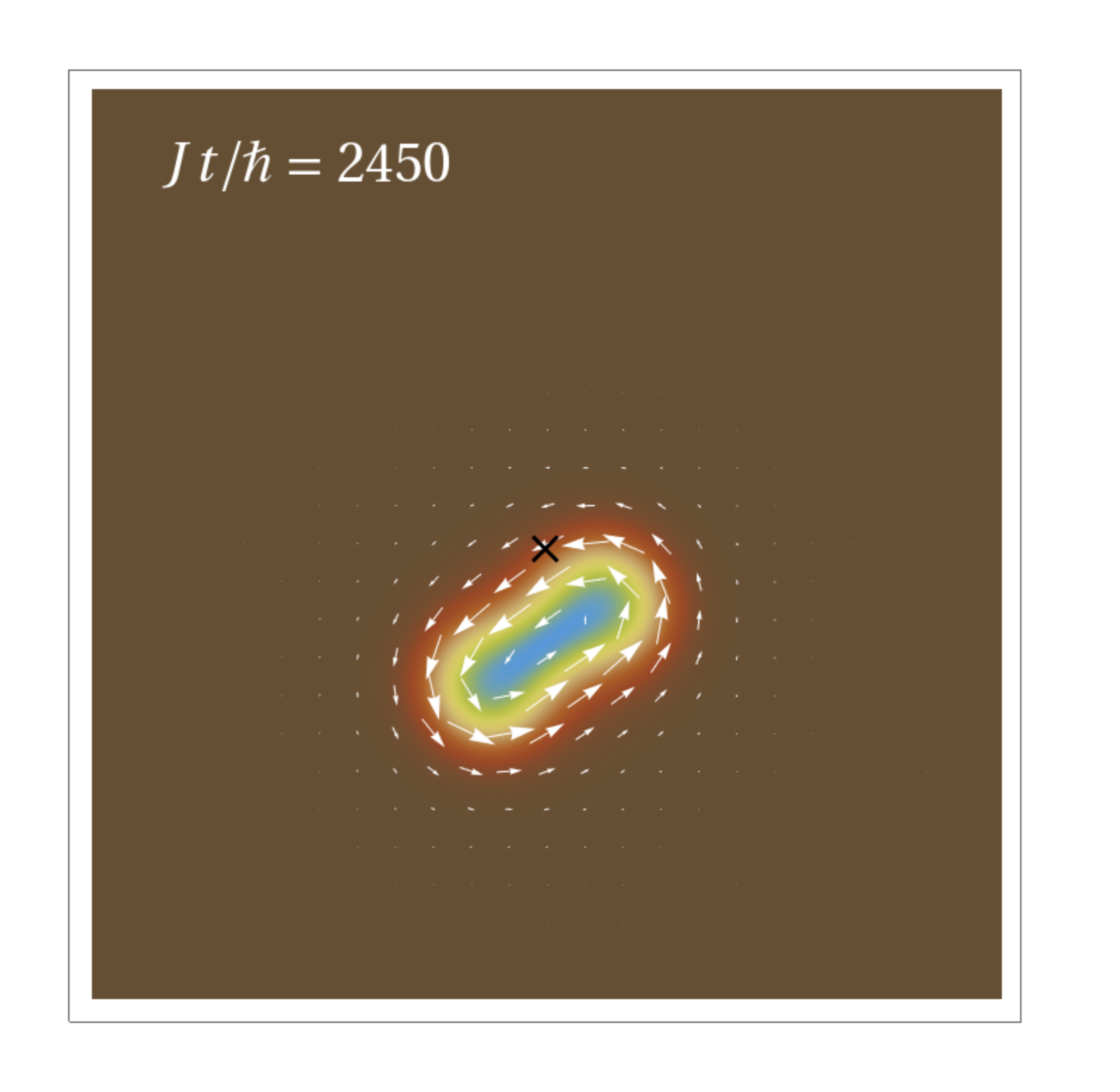}
\end{minipage}
\begin{minipage}{0.45\linewidth}
\includegraphics[width=\textwidth]{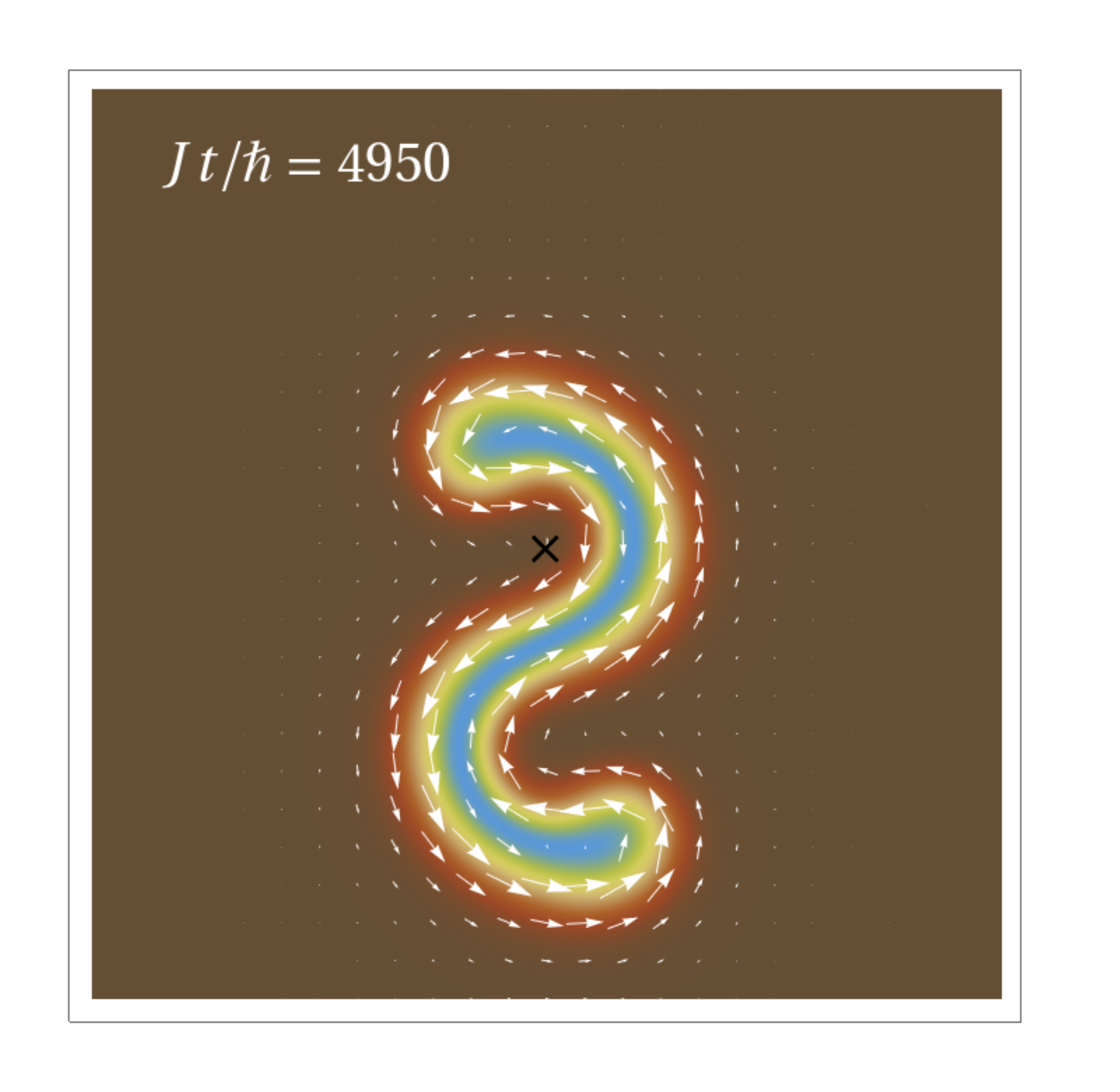}
\end{minipage}
\begin{minipage}{0.45\linewidth}
\includegraphics[width=\textwidth]{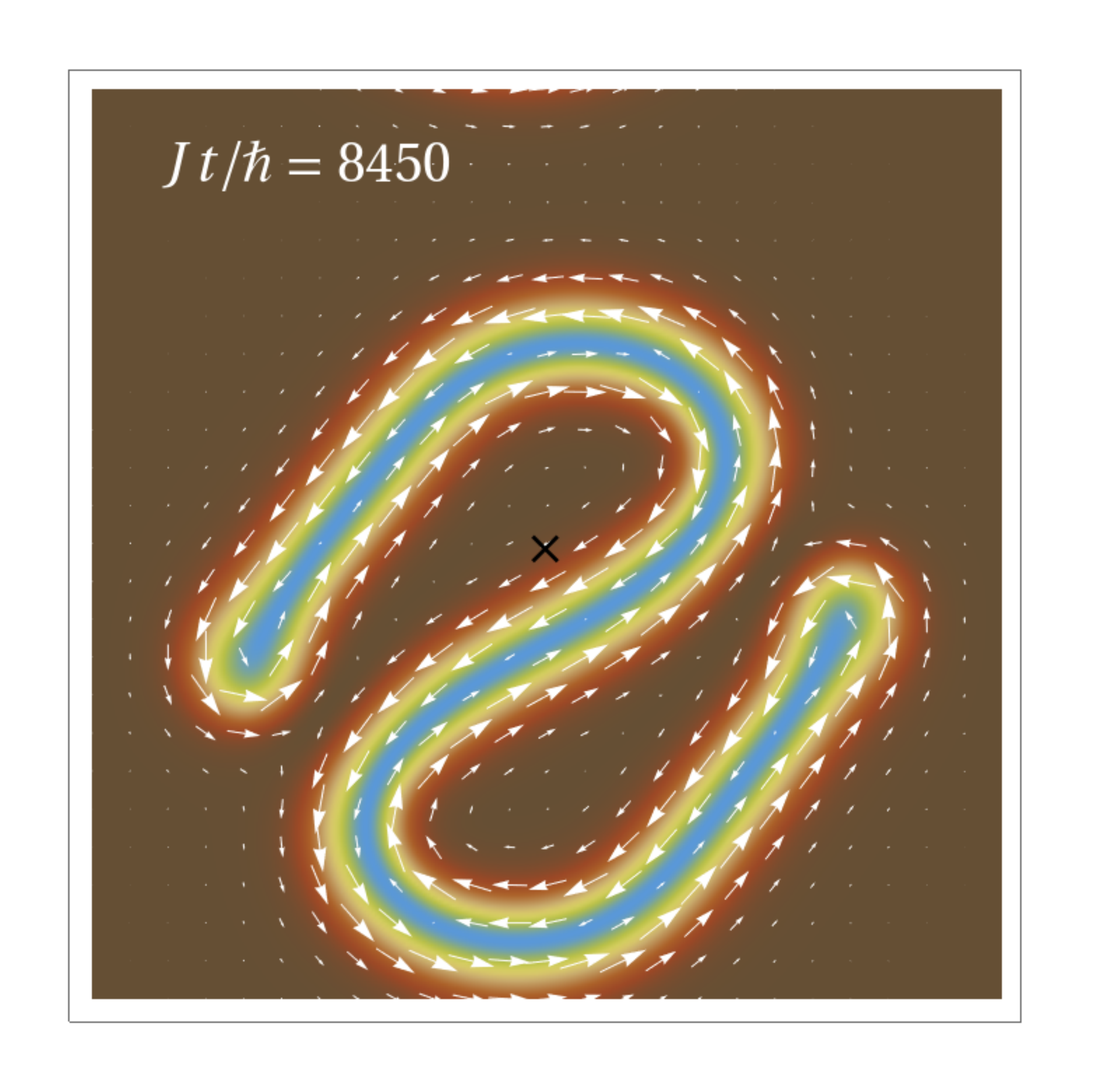}
\end{minipage}
\caption{Evolution of the spin texture obtained by numerically solving the LLG equations for $H/(JS) = -0.0095$, $A/J = 0.13$, and $E_D/J = 0.001$. The center of the defect is marked with an ``x". The skyrmion is initially relaxed to the equilibrium configuration for these material parameters, then a defect is introduced at the center of the lattice.}
\label{fig:sz-skyrmion-domain-walls}
\end{figure}

In our study of skyrmion behavior near a defect, we find several possible outcomes depending on both the material parameters and the distance of the skyrmion from the defect. At a nonzero damping the skyrmion initialized close to the defect will slowly spiral toward the defect, as is illustrated in Fig. \ref{fig:matplot-small-damping}. The range of $\Delta/a$ where this occurs corresponds to a nonzero curvature of $E(\Delta)$ in Fig. \ref{fig:energy-vs-distance}. When the magnetic field is low and the skyrmion is initialized sufficiently far from the defect, we observe that an otherwise stable skyrmion gradually deforms into a snake domain as is illustrated in Fig. \ref{fig:sz-skyrmion-domain-walls}.

It has been shown analytically that skyrmion deformation can be used to determine the phase boundary between skyrmion and domain states in a homogeneous system without defects \cite{Bogdanov1994}. The elliptical deformations of the skyrmion were used to obtain the field below which the skyrmion begins to deform into a lower-energy elongated domain. Our micromagnetic calculations show that the presence of the defect transforms the otherwise stable skyrmion into a snake domain. It is interesting to notice that close to the stability threshold the skyrmion is very sensitive to the presence of the defect. Its transformation into a snake domain occurs even when it is relatively far from the defect. 

\begin{figure}
\centering
\includegraphics[width=0.75\linewidth]{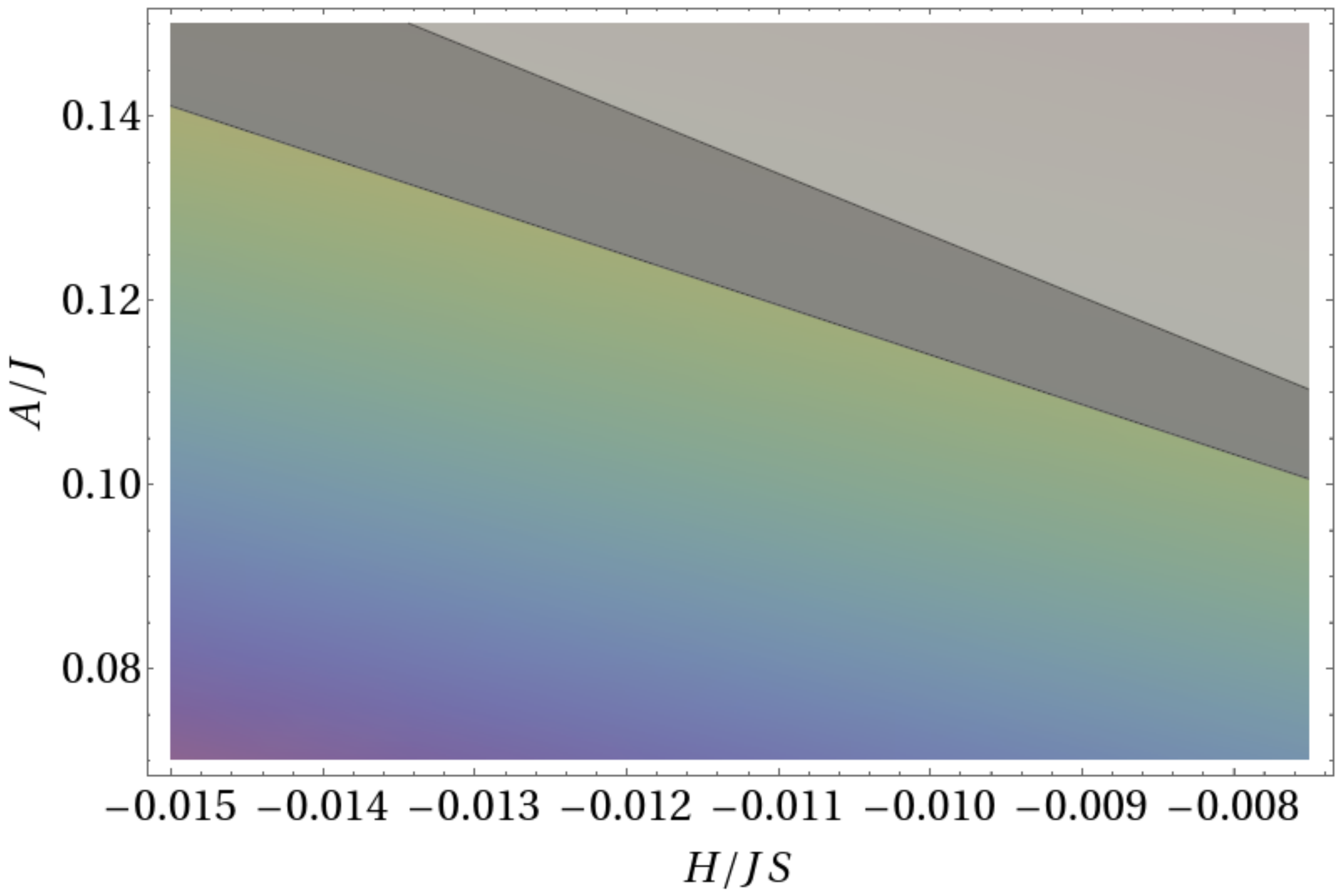}
\includegraphics[width=0.65\linewidth]{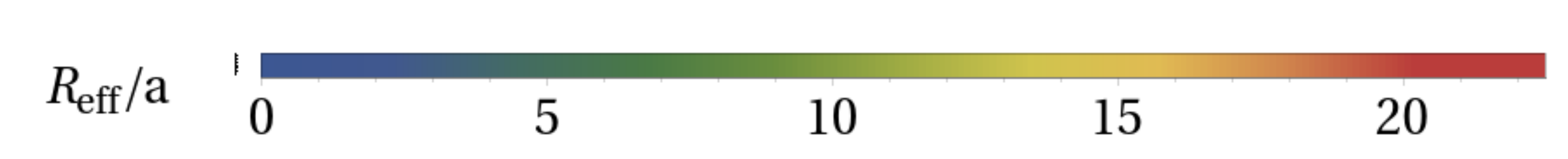}
\caption{Phases diagram of skyrmion states with and without a defect. Dark gray area represents area of the phase diagram for which the skyrmion deforms into snake domains when a defect is nearby, as shown in Fig. \ref{fig:sz-skyrmion-domain-walls}. When a defect is not present, the phase diagram continues into this region. Colors correspond to skyrmion size}
\label{fig:phase-diagram}
\end{figure}

This effect can be illustrated by the phase diagram for the skyrmion in the presence of a defect. We obtain it by initializing a BP skyrmion at a fixed distance, $\Delta/a = 28$, from the center of the lattice and evaluating the minimum energy configuration of the spins by solving the LLG equation, similar to how it was done in determining the energy versus distance in Section \ref{skyrmion-defect-interaction}. The field is then increased incrementally to obtain equilibrium skyrmion configurations for various magnetic fields and DMI values. The results are shown in Fig. \ref{fig:phase-diagram}.

\newpage

\section{Conclusion}\label{conclusion}

We have shown that an exchange-reducing defect exerts an attractive force on a nearby skyrmion. This has been done by employing micromagnetic computations that include the exchange, Zeeman, DMI, PMA, and DDI energies. The interaction energy between a skyrmion and a defect is computed as a function of the distance between the two objects. The results show that the energy of the skyrmion is lower when it centers on the defect. We also have observed skyrmion attraction to the defect in a dynamical micromagnetic computation where a skyrmion initialized near the defect spirals toward the defect. It has been compared with the results of the Thiele dynamics and good agreement has been found. Small discrepancies between the two models are attributed to the deformation of the skyrmion by the defect.

In some range of parameters, the defect induces strong deformation of the skyrmion leading to the formation of a snake-like magnetic domain. It has been observed directly through snapshots of the micromagnetic computation, as well as through the computed modification of the skyrmion phase diagram in a system containing a defect. This effect occurs at higher magnetic fields than the fields needed to transform skyrmions into labyrinth domains in a homogeneous system without defects. In a certain range of the field the skyrmion appears to be very sensitive to the presence of the defect. Even far away from the defect an otherwise stable skyrmion deforms into a snake domain. This can be used for sensing crystal defects with skyrmions. Our results indicate that the presence of defects may strongly affect application of skyrmions in memory tracks. 

We additionally show by analytical methods that skyrmion deformation should be considered when quantifying the energy of a skyrmion near a defect. We begin with the defect-free skyrmion profile of cylindrical symmetry and show that a defect generates elliptical perturbation of such a profile. These analytical results confirm skyrmion tendency to deform near a defect that has been observed numerically. 

Some useful proposals of skyrmion manipulation employ external means to move them in a material \cite{Kang2016a, Fook2016}. For applications it may also be useful to better understand the role of defects and their potential to control skyrmion position by internal means. While our work focused on the impact that a change in the exchange interaction has on a skyrmion, the modification of other energies would also be useful as additional types of skyrmion pinning  \cite{Hanneken2016, Stosic2017, Navau2018}. For example, DMI can also be controlled by defects of the crystal structure as well as through modification of the thickness of the film  \cite{Mulkers2017, Hrabec2014, Cho2015, Ma2016}. 

It has been shown that skyrmions are relatively mobile even in environments with defects \cite{Kim2017, Iwasaki2013, Sampaio2013}. Due to their topological charge magnetic skyrmions typically move in curved paths \cite{Koshibae2017}, but there may be some applications where other trajectories are needed. It remains to be seen whether skyrmion motion can be controlled by strategic placement of defects. Our computation of skyrmion interaction with a defect is a step in this direction. 

We hope that effects discussed in this paper, such as deformation and pinning of the skyrmion by a crystal defect, the depinning of the skyrmion by a current, and its transformation into a snake domain by a defect, will trigger experiments and will be observed. 

Our final comment is about a source of the local modification of the exchange other than the crystal defect. An electron or a hole in a material with low conductivity forms a polaron state characterized by a localized elastic deformation. That deformation must modify the exchange interaction in a manner similar to that from a crystal defect. Thus one should expect similar effects, with the coupling between such a polaron and a skyrmion being one of them. This would create a possibility of manipulating skyrmions with the electric fields.

\appendix

\section{Numerical Method for Micromagnetic Approach}\label{appendix-micromagnetic}\label{appendix-llg}

\begin{figure}
\centering
\includegraphics[width=0.7\linewidth]{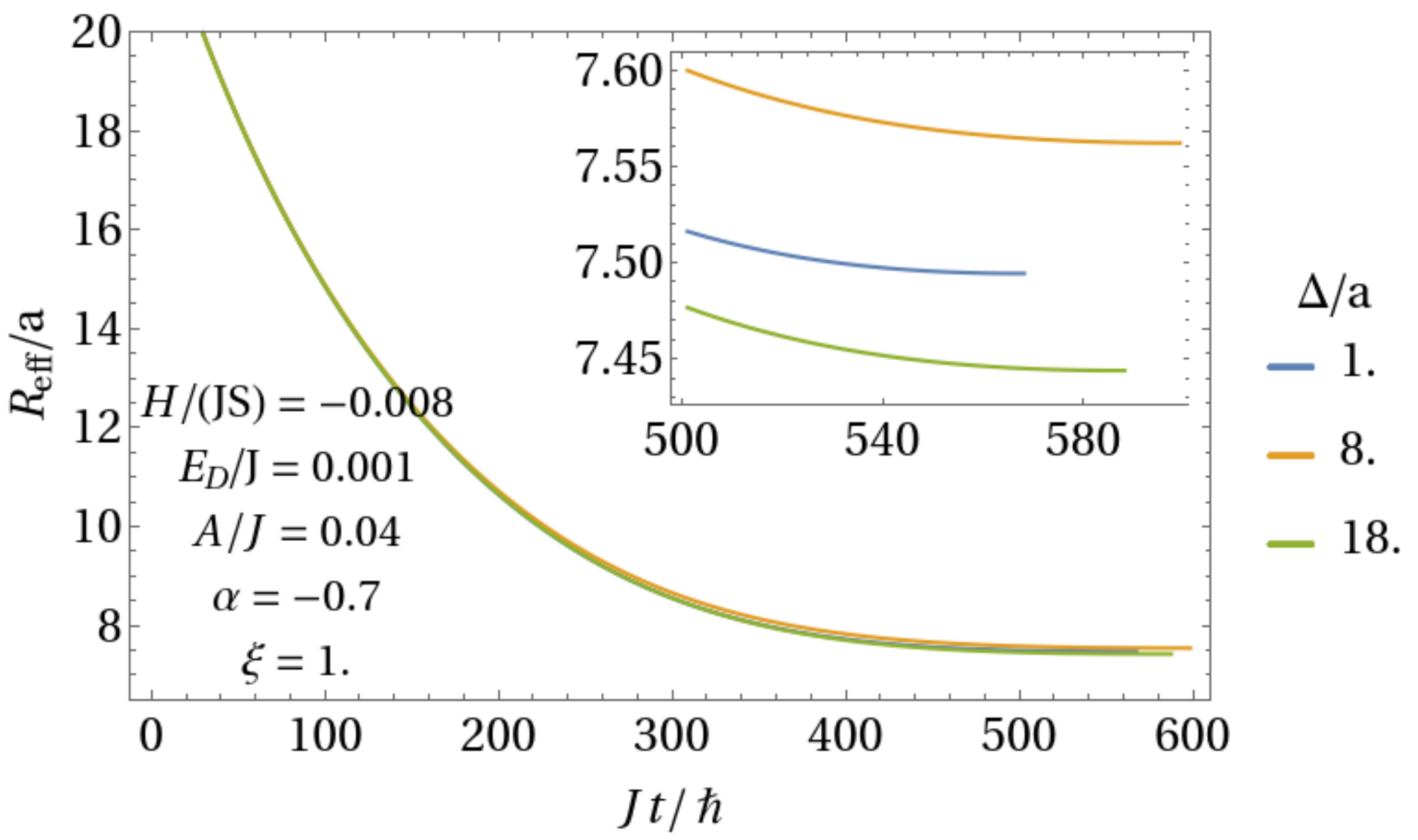}
\caption{Effective skyrmion size as a function of time in a film containing a skyrmion and a defect. The minimum energy configuration is determined when the effective size of the skyrmion converges to a tolerance of $10^{-6}$. Inset shows the effective size versus time at the end of the relaxation.}
\label{fig:size-vs-time-relaxation}
\end{figure}

We outline the numerical methods used to determine the minimum energy configuration for a skyrmion initialized near a defect. We also use these methods to compute the time evolution of the skyrmion as it rotates around the defect. We additionally elaborate on the pinning process used in fixing a skyrmion different distances from the defect.

To find the minimum energy configuration, we initialize the lattice of spins to a skyrmion-like texture, then modify each spin in a way that leads to some optimal energy. We do this by computing the time evolution of each spin using the Landau-Lifshitz-Gilbert (LLG) equation. Each spin obeys

\begin{align}\label{llg}
\dot{\mathbf{s}}_i &=  \gamma (\mathbf{s_i} \times \mathbf{H}^{eff}_i) - \lambda \, (\mathbf{s}_i \times (\mathbf{s}_i \times \mathbf{H}^{eff}_i)) \nonumber \\
&=  \gamma (\mathbf{s}_i \times \mathbf{H}^{eff}_i) + \lambda \, (\mathbf{H}^{eff}_i - \mathbf{s}_i (\mathbf{s}_i \cdot \mathbf{H}^{eff}_i)),
\end{align}

\noindent where $\mathbf{H}^{eff}_i = \partial E/ \partial  \mathbf{s}_i$ is the effective field at spin $i$, $\gamma$ is the gyromagnetic ratio which we set to $1$ in numerical work, and $\lambda$ is a damping constant. To find the minimum energy configuration, we compute spin dynamics with the LLG equation, setting $\lambda$ to $1.0$.

We begin the micromagnetic computation by initializing a Belavin-Polyakov (BP) skyrmion centered on the $y$ axis. The components of the BP skyrmion are \cite{Belavin1975}

\begin{align} \label{BP}
    s_{x} & = 2 R \frac{r(\cos\phi \cos\zeta - \sin\phi \sin\zeta)}{r^2 + R^2}, \nonumber \\
    s_y & = 2 R \frac{r(\sin\phi \cos\zeta + \cos\phi \sin\zeta)}{r^2 + R^2}, \nonumber \\
    s_z & = \frac{R^2 - r^2}{R^2 + r^2}.
\end{align}

\noindent Here $R$ is the skyrmion radius, $r$ denotes position in the $xy$ plane, and $\zeta$ is the chirality of the skyrmion. For a Bloch-type BP skyrmion, $\zeta$ = $\pi/2$. 

Pinning of the skyrmion is achieved by applying a fictitious field opposite to the external field, $H$, at the center of the skyrmion. This method was previously employed in exploring skyrmion behavior near a defect \cite{Lin2013} and was used recently in determining the interaction energy between two skyrmions \cite{Capic2020}. It has been shown in Ref. \cite{Capic2020} that pinning does not distort the skyrmion significantly from the BP texture. Here we additionally compare a pinned skyrmion to an free one of the same parameters and found that the two differ in radius by less than $0.3\%$ when a pinning field of $0.01JS$ is applied. Similar small differences in energy were observed. 

We compute skyrmion motion by including spin transfer torque in the LLG, expressed in Eq. \eqref{llg}. The additional term is $(\mathbf{j}\cdot \nabla)\mathbf{s}$, where $\mathbf{j}$ is the current density. We implement this term numerically by approximating the gradient with a first order finite difference scheme. Specifically, we add $-\sum_{\delta = \hat{x}, \hat{y} } [ j_{\delta} \mathbf{s}_{\mathbf{r}} \times ( \mathbf{s}_{\mathbf{r}+\delta} - \mathbf{s}_{\mathbf{r}-\delta} )/2  \times \mathbf{s}_{\mathbf{r}} ]$ to Eq. \eqref{llg} \cite{Liu2013}. In all computations we only apply a current in the $y$ direction.

A fourth-order Runge-Kutta numerical scheme has been used to find evolution of each spin in time. The computation stops when the effective size of the skyrmion converges to a specified tolerance, as shown in Fig. \ref{fig:size-vs-time-relaxation}. We calculate effective size using 

\begin{align}
R_{eff}^2 = a^2\frac{n-1}{2^n\pi}\sum_i{(s_{z,i}+1)^n}, 
\end{align}

\noindent which is valid for any integer $n$ \cite{Garanin2018}. In our work we set $n=2$. The tolerance for our model is set to $10^{-6}$. For practical reasons, it is also useful to set a maximum computation time so that the solver will terminate when a maximum of $t=10,000$ time steps is reached. All computations are done with periodic boundary conditions on a $256 \times 256$ lattice, following the example of Ref. \cite{Wang2010} to implement boundary conditions for DDI. Computations are performed with code written in Julia on the CUNY Lehman College 40-node computing cluster. All post processing and data analysis were done in Wolfram Mathematica. The choice of parameters is dictated by typical numbers that appear in experiments with skyrmions.

\section{Functional Approach}\label{appendix-functional}

\begin{figure}
\centering
\includegraphics[width=0.75\linewidth]{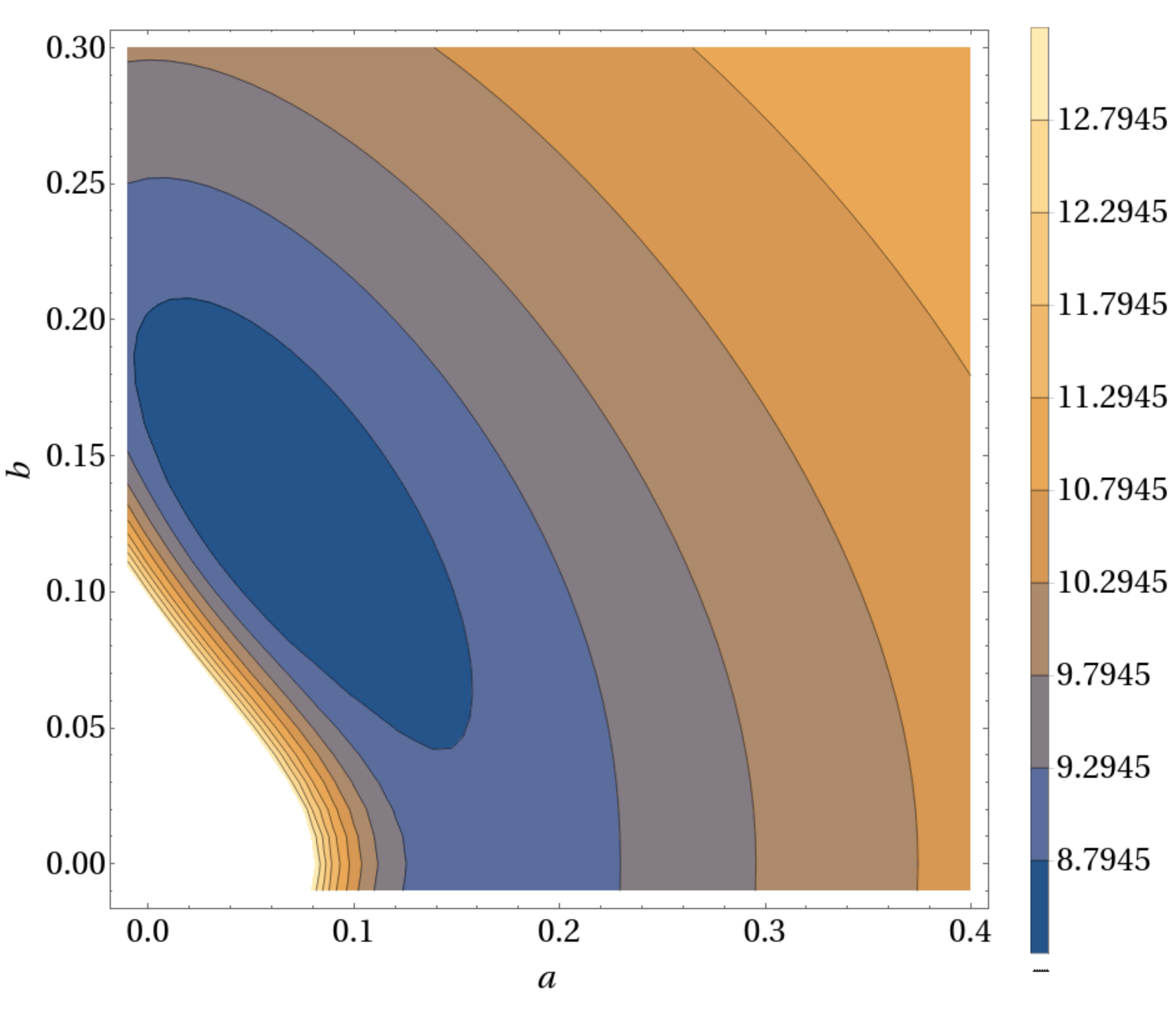}
\caption{Total energy as a function of trial function parameters, $a$ and $b$, in units of $JS^2$. A local minimum is observed. Phase space shown is for $H/(JS)=0.025$, $A/J=0.09$, and $D_z/J=0.01$. Other choices of parameters within practical limits produce similar results.}
\label{fig:energy-phase-space}
\end{figure}

We compute the function $f(r)$ that minimizes the energy of Eq. (\ref{energy-cylindrical}) by varying parameters $a$ and $b$ in the ansatz  $f = f_0(r) = \pi \exp(-a r) \text{sech}(b r)$. Since the search is for a local, and not a global, minimum, we must provide a reasonable initial guess to the optimizer. We do this by first exploring the energy phase space in terms of function parameters, then identifying a good initial guess. An example of such a phase space is shown in Fig. \ref{fig:energy-phase-space}, where a local minimum is observed. Examples of some optimized solutions are shown in Fig. \ref{fig:f-of-r-no-defect}.

\newpage

\section*{Acknowledgments}
This work has been supported by the Grant No. DE-FG02- 93ER45487 funded by the U.S. Department of Energy, Office of Science.

\section*{Data and code availability}
All codes developed in this research and the raw data are available upon request.

\section*{References}
\bibliography{skyrmion-defect-references}

\begin{thebibliography}{10}

\bibitem{Fert2017}
A.~Fert, N.~Reyren, and V.~Cros, ``{Magnetic skyrmions: Advances in physics and
  potential applications},'' {\em Nature Reviews Materials}, vol.~2, p.~17031,
  2017.

\bibitem{Kang2016b}
W.~Kang, Y.~Huang, X.~Zhang, Y.~Zhou, and W.~Zhao, ``{Skyrmion-Electronics: An
  Overview and Outlook},'' {\em Proceedings of the IEEE}, vol.~104,
  pp.~2040--2061, 2016.

\bibitem{Leonov2015}
A.~O. Leonov, T.~L. Monchesky, N.~Romming, A.~Kubetzka, A.~N. Bogdanov, and
  R.~Wiesendanger, ``{The properties of isolated chiral skyrmions in thin
  magnetic films},'' {\em New Journal of Physics}, vol.~18, p.~065003, 2016.

\bibitem{Romming2013}
N.~Romming, C.~Hanneken, M.~Menzel, J.~E. Bickel, B.~Wolter, K.~von Bergmann,
  A.~Kubetzka, and R.~Wiesendanger, ``{Writing and Deleting Single Magnetic
  Skyrmions},'' {\em Science}, vol.~341, no.~6146, pp.~636--639, 2013.

\bibitem{Yu2010}
X.~Z. Yu, Y.~Onose, N.~Kanazawa, J.~H. Park, J.~H. Han, Y.~Matsui, N.~Nagaosa,
  and Y.~Tokura, ``{Real-space observation of a two-dimensional skyrmion
  crystal},'' {\em Nature}, vol.~465, no.~7300, pp.~901--904, 2010.

\bibitem{Heinze2011}
S.~Heinze, K.~{Von Bergmann}, M.~Menzel, J.~Brede, A.~Kubetzka,
  R.~Wiesendanger, G.~Bihlmayer, and S.~Bl{\"{u}}gel, ``{Spontaneous
  atomic-scale magnetic skyrmion lattice in two dimensions},'' {\em Nature
  Physics}, vol.~7, pp.~713--718, 2011.

\bibitem{Liu2013}
Y.~H. Liu and Y.~Q. Li, ``{A mechanism to pin skyrmions in chiral magnets},''
  {\em Journal of Physics: Condensed Matter}, vol.~25, p.~076005, 2013.

\bibitem{Lin2013}
S.~Z. Lin, C.~Reichhardt, C.~D. Batista, and A.~Saxena, ``{Particle model for
  skyrmions in metallic chiral magnets: Dynamics, pinning, and creep},'' {\em
  Physical Review B}, vol.~87, p.~214419, 2013.

\bibitem{Navau2018}
C.~Navau, N.~Del-Valle, and A.~Sanchez, ``{Interaction of isolated skyrmions
  with point and linear defects},'' {\em Journal of Magnetism and Magnetic
  Materials}, vol.~465, pp.~709--715, 2018.

\bibitem{Stosic2017}
D.~Stosic, T.~B. Ludermir, and M.~V. Milo{\v{s}}evi{\'{c}}, ``{Pinning of
  magnetic skyrmions in a monolayer Co film on Pt(111): Theoretical
  characterization and exemplified utilization},'' {\em Physical Review B},
  vol.~96, p.~214403, 2017.

\bibitem{LimaFernandes2018}
I.~{Lima Fernandes}, J.~Bouaziz, S.~Bl{\"{u}}gel, and S.~Lounis,
  ``{Universality of defect-skyrmion interaction profiles},'' {\em Nature
  Communications}, vol.~9, p.~4395, 2018.

\bibitem{Hanneken2016}
C.~Hanneken, A.~Kubetzka, K.~{Von Bergmann}, and R.~Wiesendanger, ``{Pinning
  and movement of individual nanoscale magnetic skyrmions via defects},'' {\em
  New Journal of Physics}, vol.~18, p.~055009, 2016.

\bibitem{Pulecio2016}
J.~F. Pulecio, A.~Hrabec, K.~Zeissler, R.~M. White, Y.~Zhu, and C.~H. Marrows,
  ``{Hedgehog Skyrmion Bubbles in Ultrathin Films with Interfacial
  Dzyaloshinskii-Moriya Interactions},'' pp.~1--13, 2016.

\bibitem{Zeissler2017}
K.~Zeissler, M.~Mruczkiewicz, S.~Finizio, J.~Raabe, P.~M. Shepley, A.~V.
  Sadovnikov, S.~A. Nikitov, K.~Fallon, S.~McFadzean, S.~McVitie, T.~A. Moore,
  G.~Burnell, and C.~H. Marrows, ``{Pinning and hysteresis in the field
  dependent diameter evolution of skyrmions in Pt/Co/Ir superlattice stacks},''
  {\em Scientific Reports}, vol.~7, p.~15125, 2017.

\bibitem{Arjana2020}
I.~G. Arjana, I.~L. Fernandes, J.~Chico, and S.~Lounis, ``{Sub-nanoscale
  atom-by-atom crafting of skyrmion-defect interaction profiles},'' no.~111,
  2020.

\bibitem{Garanin2020}
D.~A. Garanin, E.~M. Chudnovsky, S.~Zhang, and X.~Zhang, ``{Thermal creation of
  skyrmions in ferromagnetic films with perpendicular anisotropy and
  Dzyaloshinskii-Moriya interaction},'' {\em Journal of Magnetism and Magnetic
  Materials}, vol.~493, p.~165724, 2020.

\bibitem{Salimath2019}
A.~Salimath, A.~Abbout, A.~Brataas, and A.~Manchon, ``{Current-driven skyrmion
  depinning in magnetic granular films},'' {\em Physical Review B}, vol.~99,
  no.~10, pp.~1--8, 2019.

\bibitem{Litzius2020}
``{The role of temperature and drive current in skyrmion dynamics},'' {\em
  Nature Electronics}, vol.~3, no.~1, pp.~30--36, 2020.

\bibitem{Muller2015}
J.~M{\"{u}}ller and A.~Rosch, ``{Capturing of a magnetic skyrmion with a
  hole},'' {\em Physical Review B}, vol.~91, p.~054410, 2015.

\bibitem{Psaroudaki2020}
C.~Psaroudaki and D.~Loss, ``{Quantum Depinning of a Magnetic Skyrmion},'' {\em
  Physical Review Letters}, vol.~124, p.~097202, 2020.

\bibitem{Kim2017}
J.~V. Kim and M.~W. Yoo, ``{Current-driven skyrmion dynamics in disordered
  films},'' {\em Applied Physics Letters}, vol.~110, p.~132404, 2017.

\bibitem{Buttner2018}
F.~B{\"{u}}ttner, I.~Lemesh, and G.~S. Beach, ``{Theory of isolated magnetic
  skyrmions: From fundamentals to room temperature applications},'' {\em
  Scientific Reports}, vol.~8, p.~16675, 2018.

\bibitem{Pierobon2018}
L.~Pierobon, C.~Moutafis, Y.~Li, J.~F. L{\"{o}}ffler, and M.~Charilaou,
  ``{Collective antiskyrmion-mediated phase transition and defect-induced
  melting in chiral magnetic films},'' {\em Scientific Reports}, vol.~8,
  p.~16675, 2018.

\bibitem{Capic2019}
D.~Capic, D.~A. Garanin, and E.~M. Chudnovsky, ``{Stabilty of biskyrmions in
  centrosymmetric magnetic films},'' {\em Physical Review B}, vol.~100,
  p.~014432, 2019.

\bibitem{Bogdanov1994}
A.~N. Bogdanov and A.~Hubert, ``{The Properties of Isolated Magnetic
  Vortices},'' {\em J. Magn. Magn. Mater}, vol.~138, p.~255, 1994.

\bibitem{Kang2016a}
W.~Kang, Y.~Huang, C.~Zheng, W.~Lv, N.~Lei, Y.~Zhang, X.~Zhang, Y.~Zhou, and
  W.~Zhao, ``{Voltage controlled magnetic skyrmion motion for racetrack
  memory},'' {\em Scientific Reports}, vol.~6, p.~23164, 2016.

\bibitem{Fook2016}
H.~T. Fook, W.~L. Gan, and W.~S. Lew, ``{Gateable Skyrmion Transport via
  Field-induced Potential Barrier Modulation},'' {\em Scientific Reports},
  vol.~6, p.~21099, 2016.

\bibitem{Mulkers2017}
J.~Mulkers, B.~{Van Waeyenberge}, and M.~V. Milo{\v{s}}evi{\'{c}}, ``{Effects
  of spatially engineered Dzyaloshinskii-Moriya interaction in ferromagnetic
  films},'' {\em Physical Review B}, vol.~95, p.~144401, 2017.

\bibitem{Hrabec2014}
A.~Hrabec, N.~A. Porter, A.~Wells, M.~J. Benitez, G.~Burnell, S.~McVitie,
  D.~McGrouther, T.~A. Moore, and C.~H. Marrows, ``{Measuring and tailoring the
  Dzyaloshinskii-Moriya interaction in perpendicularly magnetized thin
  films},'' {\em Physical Review B}, vol.~90, p.~020403(R), 2014.

\bibitem{Cho2015}
J.~Cho, N.~H. Kim, S.~Lee, J.~S. Kim, R.~Lavrijsen, A.~Solignac, Y.~Yin, D.~S.
  Han, N.~J. {Van Hoof}, H.~J. Swagten, B.~Koopmans, and C.~Y. You,
  ``{Thickness dependence of the interfacial Dzyaloshinskii-Moriya interaction
  in inversion symmetry broken systems},'' {\em Nature Communications}, vol.~6,
  p.~7635, 2015.

\bibitem{Ma2016}
X.~Ma, G.~Yu, X.~Li, T.~Wang, D.~Wu, K.~S. Olsson, Z.~Chu, K.~An, J.~Q. Xiao,
  K.~L. Wang, and X.~Li, ``{Interfacial control of Dzyaloshinskii-Moriya
  interaction in heavy metal/ferromagnetic metal thin film heterostructures},''
  {\em Physical Review B}, vol.~94, p.~180408(R), 2016.

\bibitem{Iwasaki2013}
J.~Iwasaki, M.~Mochizuki, and N.~Nagaosa, ``{Universal current-velocity
  relation of skyrmion motion in chiral magnets},'' {\em Nature
  Communications}, vol.~4, p.~1463, 2013.

\bibitem{Sampaio2013}
J.~Sampaio, V.~Cros, S.~Rohart, A.~Thiaville, and A.~Fert, ``{Nucleation,
  stability and current-induced motion of isolated magnetic skyrmions in
  nanostructures},'' {\em Nature Nanotechnology}, vol.~8, pp.~839--844, 2013.

\bibitem{Koshibae2017}
W.~Koshibae and N.~Nagaosa, ``{Theory of skyrmions in bilayer systems},'' {\em
  Scientific Reports}, vol.~7, p.~42645, 2017.

\bibitem{Belavin1975}
A.~A. Belavin and A.~M. Polyakov, ``{Metastable states of two-dimensional
  isotropic ferromagnets},'' {\em J. exper. theor. Phys. Letters}, vol.~22,
  p.~10, 1975.

\bibitem{Capic2020}
D.~Capic, D.~A. Garanin, and E.~M. Chudnovsky, ``{Skyrmion-Skyrmion Interaction
  in a Magnetic Film},'' {\em Journal of Physics: Condensed Matter}, vol.~32,
  p.~41, 2020.

\bibitem{Garanin2018}
D.~A. Garanin, D.~Capic, S.~Zhang, X.~Zhang, and E.~M. Chudnovsky, ``{Writing
  skyrmions with a magnetic dipole},'' {\em Journal of Applied Physics},
  vol.~124, p.~113901, 2018.

\bibitem{Wang2010}
W.~Wang, C.~Mu, B.~Zhang, Q.~Liu, J.~Wang, and D.~Xue, ``{Two-dimensional
  periodic boundary conditions for demagnetization interactions in
  micromagnetics},'' {\em Computational Materials Science}, vol.~49,
  pp.~84--87, 2010.

\end{thebibliography}
\bibliographystyle{ieeetr}

\end{document}